%% file: main.tex
\documentclass[sigconf,natbib,screen=true,review=false,anonymous=false]{acmart}

\acmSubmissionID{354}

\input{packages}
\input{meta}
\input{definitions}
\input{authors}

\copyrightyear{2022} 
\acmYear{2022} 
\setcopyright{acmlicensed}
\acmConference[WSDM '22] {Proceedings of the Fifteenth ACM International Conference on Web Search and Data Mining}{February 21--25, 2022}{Tempe, AZ, USA.}
\acmBooktitle{Proceedings of the Fifteenth ACM International Conference on Web Search and Data Mining (WSDM '22), February 21--25, 2022, Tempe, AZ, USA}
\acmPrice{15.00}
\acmDOI{10.1145/3488560.3498441} 
\acmISBN{978-1-4503-9132-0/22/02}

\begin{document}
\fancyhead{}

\title[Understanding and Mitigating the Effect of Outliers in Fair Ranking]{Understanding and Mitigating the Effect of Outliers\\ in Fair Ranking}

\renewcommand{\shortauthors}{Sarvi et al.}

\input{sections/00-abstract.tex}




\maketitle

\acresetall

\input{sections/01-intro.tex}
\input{sections/02-background.tex}
\input{sections/03-user-perception}
\input{sections/04-method.tex}

\input{sections/05-experiments.tex}

\input{sections/06-results.tex}
\input{sections/07-related-work.tex}
\input{sections/08-conclusion.tex}

\vspace*{-2mm}
\section*{Data and Code}
To facilitate reproducibility of our work, all code and parameters are shared at \url{https://github.com/arezooSarvi/OMIT_Fair_ranking}.

\vspace*{-2mm}
\begin{acks}
This research was supported by Ahold Delhaize, 
the NWO Innovational Research Incentives Scheme Vidi (016.Vidi.189.039),
and the Hybrid Intelligence Center,  a 10-year program funded by the Dutch Ministry of Education, Culture and Science through 
the Netherlands Organisation for Scientific Research, \url{https://hybrid-intelligence-centre.nl}.
All content represents the opinion of the authors, which is not necessarily shared or endorsed by their respective employers and/or sponsors.
\end{acks}

%\clearpage
\bibliographystyle{ACM-Reference-Format}
\bibliography{references}

\end{document}

%% file: packages.tex
\usepackage{amsmath}
\usepackage{amsfonts}
\usepackage{dsfont}
\usepackage{mathtools}
\usepackage{graphicx}
\usepackage{booktabs} 
\usepackage{numprint}
\usepackage{hyperref}
\npthousandsep{,}
\npdecimalsign{.}
\usepackage[inline]{enumitem}
\usepackage{balance}
\usepackage[skip=0pt]{caption}
\usepackage{multirow}
\usepackage{siunitx}
\usepackage{acronym}
\usepackage{subfig}
\usepackage{xcolor}
\usepackage{colortbl}
\usepackage{tabularx}
\usepackage{filecontents}
\usepackage{etoolbox,siunitx}
\usepackage{pgfplots}
\usepackage{caption}
\usepackage{adjustbox}
\usepackage{algorithm}
\usepackage{algpseudocode}
\usepackage{xspace}
\robustify\bfseries

%% file: meta.tex
\copyrightyear{2022}
\acmYear{2022}
\setcopyright{rightsretained}
\acmConference[WSDM '22]{Proceedings of the 15th ACM International Conference on Web Search and Data Mining}{February 21--25, 2022}{Phoenix, AZ, USA}
\acmBooktitle{The 15th ACM International Conference on Web Search and Data Mining (WSDM '22), February 21--25, 2022, Phoenix, AZ, USA}
\acmPrice{}

%% file: definitions.tex
\newcommand{\todo}[1]{\textcolor{red}{#1}}

\definecolor{darkgreen}{rgb}{0.0, 0.2, 0.13}
\definecolor{asparagus}{rgb}{0.53, 0.66, 0.42}
\definecolor{cambridgeblue}{rgb}{0.64, 0.76, 0.68}

\newcommand{\moh}[1]{\textcolor{purple}{\textbf{[#1]}}}

\newcommand{\header}[1]{\vspace{1mm}\noindent\textbf{#1.}}
\newcommand{\headernodot}[1]{\vspace{1mm}\noindent\textbf{#1}}

\newcommand{\subheader}[1]{\vspace{1mm}\indent\textit{#1.}}

 \algrenewcommand\algorithmicrequire{\textbf{Input:}}
\algrenewcommand\algorithmicensure{\textbf{Output:}}

\AtBeginDocument{%
  \providecommand\BibTeX{{%
    \normalfont B\kern-0.5em{\scshape i\kern-0.25em b}\kern-0.8em\TeX}}}

\newtheorem{corollary }{Corollary}

\newcommand{\Exposure}{\operatorname{Exposure}}

\newcommand{\ObservableFeature}{g}
\newcommand{\OutlierDetection}{\operatorname{M}}
\newcommand{\ranking}{\sigma}
\newcommand{\DegreeOfOutlierness}[2]{\OutlierDetection(\ObservableFeature(#1)|#2)}
\newcommand{\OutlierVector}{\mathbf{o}}
\newcommand{\TopKOnes}{\mathbf{h}}
\newcommand{\StochasticRanking}{\pi}
\newcommand{\topk}{top-$k$\xspace}
\newcommand{\baseline}{RO\xspace}
\newcommand{\contextSize}{n\xspace}

\newcommand{\OurMethod}{OMIT\xspace}

\acrodef{LTR}{Learning to Rank\xspace}
\acrodef{MO}{multi-objective\xspace}
\acrodef{MRP}{marginal rank probability\xspace}
\acrodef{nDCG}{Normalized Discounted Cumulative Gain\xspace}

%% file: authors.tex
\author{%
Fatemeh Sarvi
}
\affiliation{%
  \institution{
  AIRLab, University of Amsterdam
  \city{Amsterdam}
  \country{The Netherlands}%
  }
}  
\email{f.sarvi@uva.nl}

\author{%
Maria Heuss
}
\affiliation{%
  \institution{
  University of Amsterdam
  \city{Amsterdam}
  \country{The Netherlands}%
  }
}  
\email{m.c.heuss@uva.nl}

\author{%
Mohammad Aliannejadi
}
\affiliation{%
  \institution{
  University of Amsterdam
  \city{Amsterdam}
  \country{The Netherlands}%
  }
}  
\email{m.aliannejadi@uva.nl}

\author{%
Sebastian Schelter
}
\affiliation{%
  \institution{
  University of Amsterdam
  \city{Amsterdam}
  \country{The Netherlands}%
  }
}
\email{s.schelter@uva.nl}

\author{%
Maarten de Rijke%
}
\affiliation{%
  \institution{
  University of Amsterdam
  \city{Amsterdam}
  \country{The Netherlands}%
  }  
}  
\email{m.derijke@uva.nl}

%% file: sections/00-abstract.tex
% !TEX root = ../main.tex

\begin{abstract}
Traditional ranking systems are expected to sort items in the order of their relevance and thereby maximize their utility. 
In fair ranking, utility is complemented with fairness as an optimization goal.
Recent work on fair ranking focuses on developing algorithms to optimize for fairness, given position-based exposure. 
In contrast, we identify the potential of outliers in a ranking to influence exposure and thereby negatively impact fairness.
An outlier in a list of items can alter the examination probabilities, which can lead to different distributions of attention, compared to position-based exposure.  
%
%\mdr{of what}
%We claim that ignoring inter-item dependencies in a list can itself become a source of unfairness. 
%Exposure must be a true estimate of the attention each item receives. 
%We define \emph{outliers} as task-specific, property-dependant items that deviate from the rest in the ranking. We hypothesize that having an outlier among items can alter examination probabilities which can lead to different values, compared to position-based exposure.  
%For example, in e-commerce search, platforms assign multiple tags to items, such as special offers and promotions. If only one item in a result page has a promotion tag, it can be considered as an outlier. 
We formalize outlierness in a ranking, show that outliers are present in realistic datasets, and present the results of an eye-tracking study, showing that users' scanning order and the exposure of items are influenced by the presence of outliers. 
We then introduce  \OurMethod{}, a method for fair ranking in the presence of outliers. Given an outlier detection method, \OurMethod{} improves fair allocation of exposure by suppressing outliers in the top-$k$ ranking. 
%Since the notion of outlierness is application-specific, we show how any outlier detection method can be adopted in this framework.
%We study the effect of outlierness optimization by providing empirical results 
Using an academic search dataset, we show that outlierness optimization leads to a fairer policy that displays fewer outliers in the top-$k$, while maintaining a reasonable trade-off between fairness and utility. 
\end{abstract}

%% file: sections/01-intro.tex
% !TEX root = ../main.tex

\section{Introduction}
\label{sec:intro}

The primary goal of a ranker as used in a search engine or recommender system is to optimize the list in order to satisfy user needs by sorting items in their order of relevance to the query~\citep{manning2008introduction}. 
Recently there has been a growing concern about the unfairness towards minority groups caused by this simplistic assumption~\citep{baeza2018bias,biega2018equity}.
Several studies have proposed approaches to achieve fair ranking policies.
%, which can be categorized as \emph{probability-based} and \emph{exposure-based}~\citep{zehlike2021fairness}. 
The goal is to ensure that the protected groups receive a predefined share of visibility. 
%Probability-based fairness fulfills this goal by finding how probable it is that a fair ranking is generated as a result of a fair process using statistical significance tests~\citep{yang2017measuring, zehlike2017fa, asudeh2019designing, celis2020interventions, celis2017ranking, geyik2019fairness, stoyanovich2018online, yang2017measuring, zehlike2017fa}. 
Exposure-based methods~\citep{biega2018equity, mehrotra2018towards, morik2020controlling, sapiezynski2019quantifying, singh2018fairness, singh2019policy} quantify the expected amount of attention each individual or group of items receives from users\if0 in a given ranking policy\fi, where attention is typically related to the item's position and based on the observation that users are more likely to click on items presented at higher positions~\citep{baeza2018bias, joachims2005accurately}. 

But item position is not the only factor that affects exposure~\citep{chuklin-2015-click}. Inter-item dependencies also play a key role~\citep{borisov-2016-neural}. E.g., consider a user who is trying to buy a phone. When searching on an e-commerce platform, if an item in the list is on promotion and has a ``Best Seller'' badge, this can be distracting so that it gets more attention from the user, irrespective of its position in the list; the item would stand out even more if it is the only one with this feature.

We hypothesize that inter-item dependencies have an effect on examination probability and exposure of items. We focus on the case of having an \emph{outlier} in the ranking and aim to understand and address its effect on user behavior. We hypothesize that exposure received by an item is influenced by the existence of an outlier in the list, and assume that this effect should be considered while allocating exposure to protected groups in a fair ranking approach.
%In this work, we take the first steps towards understanding the effect that outliers have on item exposure. 
%We hypothesize that when an item stands out in a list, the exposure it gets is not proportional to its position. %This can be a source of unfairness that either affects the item itself, or other items in the list.

We define \emph{outliers} as items that observably deviate from the rest. The properties and method with which we identify outliers in a set of items are dependent on the task. The properties are observable item features that can be presentational in nature or correspond to ranking features used \if0by the search engine\fi to produce the ranked list. E.g., in the e-commerce search example, if only one item in a result page has a ``Best Seller'' tag, it is an outlier based on this presentational feature. 
%As an example of outliers based on perceived usefulness, consider a scholarly search result page with a paper on one of the top positions with far greater number of citations that distinguishes it from the rest. Users are likely to interpret the high level of citations as more useful, which can turn this item to an outlier.   

To begin, we perform an \textbf{exploratory analysis on the TREC Fair Ranking dataset}. We observe that a large number of outliers exist in the rankings, where we use multiple outlier detection techniques to identify outliers based on the papers' citations as they can make an item more attractive and catchy than others. 

Next, we perform an \textbf{eye tracking study}, where we measure the attention that each item in a ranked list gets through eye tracking, so as to show that users can actually perceive outliers in rankings. We find that attention is more focused on outlier items. The scanning order and exposure received by each item may be influenced by the existence of  outliers.
% the following sentence is no longer useful after we've said that we have run an experiment that supports the point
%We assume that outlierness of an item affects the examination behavior and thus can be considered as a type of bias.
Unlike other types of bias studied in search and recommendation~\citep{joachims2005accurately, o2006modeling, yue2010beyond,wang2016learning,oosterhuis-2021-unifying}, our eye-tracking study reveals that outlierness comes from inter-item dependencies. It affects the examination propensities for items around the outlier in a way that is less dependent on the position  and based on relationships between items presented together. 
%Interactions between items can be used to improve ranking~\citep{pobrotyn2020context}, but this phenomenon is not well-explored in the context of fair ranking. 
%``Attractiveness'' in presentation bias~\citep{yue2010beyond} is the closest concept to what we study in this work. 
However, 
%attractiveness as proposed by~\citet{yue2010beyond} 
it is translated to bolded keyword matches in the title and abstracts, which can be calculated for each item separately, independent of its neighbors. Attractiveness does not alter the examination model based on position bias, and only results in relatively more clicks on items when they are presented with more bolded matched keywords~\citep{yue2010beyond}. 
%\todo{does this paragraph breaks the story?}
 
%Our eye tracking study suggests that outliers influence exposure of items in a way that is not purely dependent on the position. 
While allocating fair exposure to protected items or groups in a fair ranking solution, we should account for the effect of outliers. 
%In this work we take the first steps towards studying this phenomenon and leave accurately estimating item exposure in the presence of outliers as future work. As a first attempt to consider this effect, and based on our analysis of the TREC dataset and our eye tracking study, 
We \textbf{propose an approach to account for the existence of outliers in rankings without damaging the utility or fairness of the ranking} by mitigating outlierness, called \OurMethod{}.
\OurMethod{} jointly optimizes
\begin{enumerate*}
\item user utility, 
\item item fairness, and 
\item fewer outliers in top-$k$ positions
\end{enumerate*}
as a convex optimization problem that can be solved optimally through linear programming. 
Via its solution, we derive a stochastic ranking policy using Birkhoff-von Neumann (BvN) decomposition~\citep{birkhoff-1940-lattice}.
% inspired by~\citet{singh2018fairness}.
\OurMethod{} reduces the number of outliers at top-10 positions on the TREC 2020 dataset by 80.66\%, while maintaining the NDCG@10, compared to a state-of-the-art ranking baseline.
 % \moh{are you sure it is convex? if we cannot prove it we better not mention it}
 
The main contributions of this work are as follows: 
\begin{enumerate*}[leftmargin=*,nosep]
\item we introduce, study and formalize the problem of outlierness in ranking and its effects on exposure distribution and fairness;
\item we run an extensive eye-tracking user study in two search domains to support our hypothesis about the existence of an effect of outliers on items' exposure; 
\item inspired by our analysis, we propose \OurMethod{}, an efficient approach that mitigates the outlierness effect on 
fairness;
%\moh{this can be misleading. I suggest to be very careful about the usage of exposure in the paper, not to give the impression that we are modeling exposure in this work.}  
\item  we compare \OurMethod{} to competitive baselines on two TREC datasets in terms of fairness, outlierness, and utility; \OurMethod{} is able to remove outliers while balancing utility and fairness; and
\item we make the data from our eye-tracking study plus the code that implements our baselines and \OurMethod{} publicly available.\end{enumerate*} 

%% file: sections/02-background.tex
\vspace*{-2mm}
\section{Background}
\label{sec:background}

\noindent\textbf{Exposure and utility.}
Consider a single query $q$,  that we will often leave out for notational simplicity,  for which we want to rank a set of documents $\mathcal{D} = \{d_1,d_2,\dots,d_N\}$.  Suppose we are given document utilities $\mathbf{u} \in \mathbb{R}^N$,  where $u_i$ is a proxy for the relevance of document $d_i$ for $q$.  
Let $\mathbf{v} \in \mathbb{R}^N$, be the {\em attention vector},  where $v_j$ denotes how much attention a document gets at position $j$,  and which is decreasing with the position.  This vector encodes the assumed position bias, e.g., $v_j = 1 / \log(1 + j)$.   

We require a probabilistic ranking in the form of a doubly stochastic document-position matrix $\mathbf{P} \in [0,1]^{N \times N}$ where $P_{ij}$ denotes the probability of putting document $d_i$ at position $j$.  Such a matrix can be decomposed into a convex combination of permutation matrices, which allows us to sample a concrete ranking~\citep{singh2018fairness}. 

The {\em exposure} of a document $d_i$ under ranking $\mathbf{P}$ denotes the expected attention that this document will get.  Using the position based attention vector $v$, this can be modeled as a function of the ranking and position bias: $\Exposure(d_i|\mathbf{P}) = \sum_{j=1}^N P_{ij} \, v_j$. 
 
The {\em expected utility} $U$ of a ranking $\mathbf{P}$ is the sum of the documents' utilities weighted by the exposure given to them by $\mathbf{P}$:
\begin{equation}
U(\mathbf{P}) = \sum_{i=1}^N u_i \, \Exposure(d_i|\mathbf{P})= \sum_{i=1}^N \sum_{j=1}^N u_i \, P_{ij} \, v_j = \mathbf{u}^T \mathbf{P} \mathbf{v}.
\end{equation}
Without fairness considerations,  a utility-maximizing ranking can be found by sorting the documents in descending order of utility. 
 
\header{Group fairness}
Suppose now that the documents $\mathcal{D}$ can be partitioned into two disjoint sets $\mathcal{D}_{dis}$ and $\mathcal{D}_{priv}$, where documents in $\mathcal{D}_{dis}$ belong to a historically disadvantaged group (e.g., publications from not so well established institutes), and those in $\mathcal{D}_{priv}$ belong to the privileged group (e.g., publications from well-established institutes).  
We want to ensure a certain notion of fairness in the ranking. We want to avoid {\em disparate treatment} of the different groups.  
We use the {\em disparate treatment ratio}~\citep{singh2018fairness}, which measures how unequal the exposure given to the disadvantaged group (in relation to the corresponding utility of the disadvantaged group) is compared to the corresponding ratio of the privileged group, as: 
\begin{equation}
\label{eq:dtr}
\begin{split}
\mathrm{dTR}&(\mathcal{D}_{dis}, \mathcal{D}_{priv} | \mathbf{P}) = \\
&\frac{
	\sum_{d_i \in \mathcal{D}_{dis}} \Exposure(d_i|\mathbf{P}) / \sum_{d_i \in \mathcal{D}_{dis}} u_i
}{
	\sum_{d_p \in \mathcal{D}_{priv}} \Exposure(d_p|\mathbf{P}) / \sum_{d_p \in \mathcal{D}_{priv}} u_p
}.
\end{split}
\end{equation}
Note that dTR is 1 if the groups are treated \emph{fairly} and smaller than 1 if the ranking is unfair towards the disadvantaged group. 
We often encounter disparate treatment when only optimizing for the expected utility of a ranking~\cite{singh2018fairness,baeza2018bias,biega2018equity}. 
We can find a utility maximizing ranking $\mathbf{P}$ that avoids disparate treatment by solving the following optimization problem~\citep{singh2018fairness}:
\begin{equation}
\label{convex_optimization}
\begin{aligned} 
\mathbf{P} =  \operatorname{arg\,max}_{\mathbf{P}} &  \mathbf{u}^\top \mathbf{P} \mathbf{v}  & (\textit{expected utility}) \\ 
\text{such that }		& \mathds{1}^\top \mathbf{P}=\mathds{1}^\top & (\textit{row stochasticity}) \\
					& \mathbf{P} \mathds{1}=\mathds{1} &  (\textit{column stochasticity}) \\
					& 0 \leq \mathbf{P}_{ij} \leq 1 & (\textit{valid probabilities}) \\
					& \mathbf{f}^\top \mathbf{P} \mathbf{v} = 0,  & (\textit{dTR constraint})
\end{aligned}
\end{equation} 
where $\mathds{1}$ denotes a vector and $\mathbf{f}$ is the vector constructed to encode the avoidance of disparate treatment, with
\begin{equation}
f_i = 
\frac{
	\mathds{1}_{d_i \in \mathcal{D}_{dis}}
}{
	|\mathcal{D}_{dis}| \, \sum_{d_s \in \mathcal{D}_{dis}} u_s 
} -
\frac{
	\mathds{1}_{d_i \in \mathcal{D}_{priv}}
}{
	|\mathcal{D}_{priv}| \, \sum_{d_p \in \mathcal{D}_{priv}} u_p
},
\end{equation}
where $\mathds{1}_{d_i \in \mathcal{D}_{dis}}=1$ if document $d_i$ is in the disadvantaged group and $0$ otherwise (and analogously for $\mathds{1}_{d_i \in \mathcal{D}_{priv}}$)~\citep{singh2018fairness}.

%\subsection{Degrees of Outlierness}
\header{Degrees of outlierness}
\label{sec:background:degrees_of_outlierness}
Outliers are items that deviate from the rest of the data~\citep{wen-2006-ranking}. They can be interesting observations or suspicious anomalies. Either way, they are considered noise that can affect the statistical analysis.
We describe three outlier detection methods that we will use later in the paper.  Let $x=\{x_1,\ldots, x_n \, | \,  x_i \in \mathbb{R}\}$ be a set of values for which we want to identify outliers. 
\if0
%The problem of outlier detection is a fundamental issue in data mining, specifically in fraud detection, network intrusion, medical problems, and network monitoring.
Most outlier detection methods are designed to deal with big datasets. However, since in our case the number of items retrieved in response to a query is limited, we choose MAD, a heuristic that has been shown to work well for small sets. We also experiment with Median K-Nearest Neighbor, a clustering-based outlier detection method. Outlier detection methods should be able to handle the big dimensionality of the data in cases where we represent items with high dimensional feature vectors. To this end, we also consider COPOD~\citep{li2020copod}, which is a deterministic and efficient model and an ideal choice for high dimensionality data.
%In the rest of this section we outline the three outlier detection methods we chose to use in the experiments, and describe how each defines outliers.
\fi

\subheader{Median Absolute Deviation (MAD)}
Although it is common practice to use Z-scores to identify possible outliers, this can be misleading (particularly for small sample sizes) due to the fact that the maximum Z-score is at most $(n-1)/\sqrt{n}$. \citet{iglewicz1993detect} recommend using the modified Z-score:
$M_i=0.6745(x_i-\tilde{x})/\mathit{MAD}$,
where $\mathit{MAD}$ is the median absolute deviation and $\tilde{x}$ is the median of $x$. These authors recommend that modified Z-scores with an absolute value of greater than 3.5 be labeled as potential outliers.

\subheader{Median K-Nearest Neighbor (MedKNN)} This model~\citep{angiulli2002fast} uses the K-Nearest Neighbor algorithm to define a distance-based outlier detection method. For each point $x_i$ we have a value $w_k(x_i)$ as the weight calculated from the $k$ nearest neighbors; 
%$w_k(x_i)$ is then used to rank the points of the dataset. 
outliers are the points with the largest values of $w_k$.
%Different versions of this algorithm can be used, based on how $w_k$ is being calculated. 
We use Median K-Nearest Neighbor, which computes $w_k(x_i)$ as the \emph{median} distance of $x_i$ to its $k$ neighbors. To find the $k$ nearest neighbors, the method linearizes the search space and uses \emph{Hilbert space-filling curve} to search efficiently; the method scales linearly in the dimensionality and the size of the data.

\subheader{Copula-Based Outlier Detection (COPOD)}
% formulation from li2020copod 
%now that in appendix I can add all the formulations, but it would be just copying from li2020copod 
%COPOD~\cite{li2020copod}, is a novel outlier detection algorithm based on estimating tail probabilities using empirical copula. In probability theory copulas are functions that explain the dependancy of two or more random variables and contain information about how they are correlated. 
%COPOD uses empirical cumulative distribution functions (ECDFs) to compute tail probabilities (see equation\ref{ecdf}). These tail probabilities estimate the probability of observing a point at least as extreme as $x_i$ for each data point $x_i$. If $x_i$ is an outlier the probability of observing a point as extreme as $x_i$ should be small and it means that this point has rare occurrence. empirical CDFs can be computed using the following equation:
COPOD~\citep{li2020copod} is a novel outlier detection method based on estimating tail probabilities using empirical copula. COPOD uses empirical cumulative distribution functions (ECDFs) to compute tail probabilities. These tail probabilities estimate the probability of observing a point at least as extreme as $x_i$ for each data point. If $x_i$ is an outlier, the probability of observing a point as extreme as $x_i$ should be small. It means that this point has a rare occurrence. This method is deterministic and efficient, and scalable to high dimensionality data.

%% file: sections/03-user-perception.tex
% !TEX root = ../main.tex

\vspace*{-2mm}
\section{Outliers in Ranking}
\label{sec:outliers}

A common assumption \if0in the literature on fair ranking\fi is that exposure is a function of position~\citep{biega2018equity, singh2018fairness, morik2020controlling, mehrotra2018towards, wang2020fairness}. 
We argue that this assumption holds only if ranked items can be deemed similar, meaning that no item is perceived as an outlier. 
Below we introduce outliers in the context of ranking.
We then determine that outliers are present in rankings in realistic datasets.
We also report on an eye-tracking study that shows that the presence of outliers in ranked lists impacts user behavior.

\header{A definition of outliers in ranking}
%\subsection{A Definition of Outliers in Ranking}
\label{sec:outliers:definition}
%As noted in Section~\ref{sec:background}, outliers are exceptional items that deviate from the rest of the data~\citep{wen-2006-ranking}.
For a ranked list, we define outliers as items that stand out among the window of items that the user can see at once, drawing the user's attention. 
Outlier items often have (visible) characteristics that distinguish them from their neighbors.  
E.g., consider Fig.~\ref{fig:eyetracking-smartphone-outlier}, which shows a result page for the query ``smart phone''. The result page view consists of 6 products, each presented with characteristics such as title, image, and price. The item at position three deviates from the other items in terms of several visual characteristics; it has more details, some promotive tags, and bold keywords. Other features, such as more positive reviews, or a higher price, may also influence the user's perceived relevancy. In this example, the third item can be considered as an outlier according to such visual characteristics.

 \begin{figure}[t]
    \centering
    \includegraphics[width=.62\linewidth]{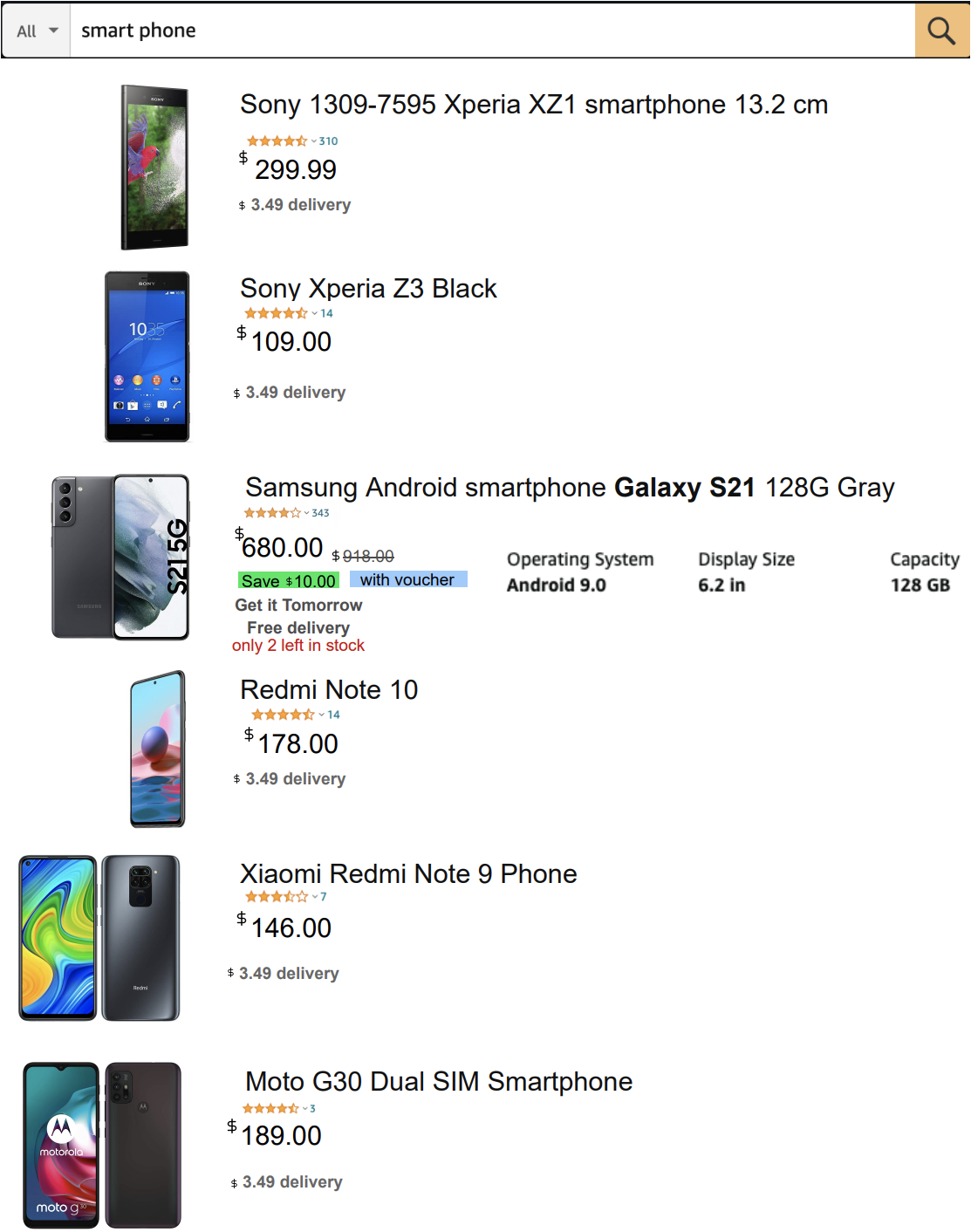}          
    \caption{E-commerce example used in our eye-tracking user study. A result page with one outlier at position 3, identified by more descriptive fields, higher price, and colored tags.}
    \label{fig:eyetracking-smartphone-outlier}
\end{figure}    

Formally, we define outliers in ranking as follows.  Consider a ranked list of $N$ items in $\mathcal{D} = \{d_1,d_2,\dots,d_N\}$,  that has been produced in response to a query. We call an observable characteristic of an item $d$ in a ranked list an \emph{observable item feature}. These features can be purely presentational in nature, like the bold keywords in Fig.~\ref{fig:eyetracking-smartphone-outlier}, or correspond to ranking features used by the search engine to produce the ranked list, e.g., the average user rating.  
\if0
Now let $\ObservableFeature$ be an observable item feature, and $\OutlierDetection$ be one of the outlier detection methods discussed in Section~\ref{sec:background:degrees_of_outlierness}. 
Note that outlier detection methods typically generate scores for each item in the set, and then compare these scores to a predefined threshold to determine whether an item is an outlier. \todo{keep, remove or put in the footnote?}
\fi

\begin{definition}[Degree of outlierness] 
Let $\ObservableFeature$ be an observable item feature, and $\OutlierDetection$ be one of the outlier detection methods discussed in Section~\ref{sec:background:degrees_of_outlierness}.
The \emph{degree of outlierness} of an item $d_i$  in the ranked list $[d_1, \ldots, d_N]$ is the value calculated by $\OutlierDetection$ for $\ObservableFeature(d_i)$ in the context of $\{\ObservableFeature(d_1), \ldots \ObservableFeature(d_N)\}$, that determines how much $\ObservableFeature(d_i)$ differs from the other elements of the set.  We write $\DegreeOfOutlierness{d_i}{\mathcal{D}}))$  for this value.
\end{definition}

 \begin{definition}[Outliers in ranking]
 We say that according to $\OutlierDetection$, item $d_i$ is an \emph{outlier in the ranked list $[d_1, \ldots, d_N]$ for feature $\ObservableFeature$}, if $\OutlierDetection$ identifies $\ObservableFeature(d_i)$ as an outlier in the set $\{\ObservableFeature(d_1), \ldots, \ObservableFeature(d_N)\}$. 
\end{definition}
\noindent%
Note that detecting an item as an outlier in a ranking depends on the context in which we see the item. Throughout the paper, we consider the full ranked list of items as the context in which we detect outliers. In Section~\ref{sec:results} we study varying sizes for the context.
 
Moreover, it is possible to use multiple observable features to detect the outliers. For example, we can consider image size as $\ObservableFeature_1$ and price as $\ObservableFeature_2$, and then use any combination of these two feature values to present item $d_i$.

Below, when we refer to an item $d$ being an outlier in a given ranked list, we assume that it is clear from the context what outlier detection method $\OutlierDetection$ and observable item feature $\ObservableFeature$ are being used. 

\headernodot{Do outliers in ranking exist?}
%\subsection{Do Outliers in Ranking Exist?}
\label{sec:outliers:outliers_in_TREC_data}
To determine whether outliers are present in rankings in datasets, we take a retrieval test collection, compute feature values for one of the (potentially observable) rankings features appropriate for the collection, and determine whether there are outliers among the top-20 documents returned for the test queries (using ListNet as the ranker, see Section~\ref{sec:experiments}).
For the experiments in Section~\ref{sec:results} we use the academic search dataset provided by the TREC Fair Ranking track.\footnote{\url{https://fair-trec.github.io/}\label{footnote:FR}} 
It contains information about papers and authors extracted from the Semantic Scholar Open Corpus.\footnote{\url{http://api.semanticscholar.org/corpus/}} 
It comes with queries and relevance judgments;
see Table~\ref{data_table} for some descriptive statistics.

\begin{table}[t]
   \caption{Descriptive statistics of TREC Fair Ranking Track 2020 data.}
   \label{data_table}
   \centering
   \begin{tabular}{lrr}
     \toprule
        &  Train  &  Test \\
      \midrule
      \#queries & 200 &  200 \\
%      \#unique authors	& 16,499&17,571 \\
      \#unique papers & 4,649 & 4,693  \\
      \% of clicks in product pages &0.169&0.170 \\
      \bottomrule
   \end{tabular}
\end{table}

We used the number of citations as observable feature $\ObservableFeature$ for this dataset as they can make an item more attractive than others (when reported). 
In the remainder of the section, we report the analysis only on the TREC 2020 data, as we observed similar trends in both datasets.
Fig.~\ref{fig:citations-in-sessions} shows the mean, maximum, and minimum of papers' citation counts for all search sessions in the data. There is a high variance between mean and maximum citations, which implies that the data is outlier-prone based on this feature. We plot outlier counts for each position in the top-20. Fig.~\ref{fig:outliers_in_sessions} depicts the number of relevant and non-relevant outliers detected by the outlier detection methods introduced in Section~\ref{sec:background} at different positions. The stacked bars show that in spite of the attractiveness of outliers, most of these items are irrelevant, judging by the click data. In total, $88.5\%$, $89.8\%$, and $90.1\%$ of the outliers are irrelevant when MedKNN, COPOD, and MAD are employed as the outlier detection method, respectively. 
The average percentage of irrelevant documents in the top 20 positions in the dataset is $83.3\%$. 
This suggests that by pushing  these items to lower positions we can improve the degree of outlierness without jeopardizing utility.

\headernodot{Do users perceive outliers in the ranking?}
Outliers are present in realistic datasets, but do they impact user behavior?
Prior studies stress the importance of relationships between ranked items~\citep{pobrotyn2020context}, but 
%there is little work that studies the effects of item-relationships on item exposure. It 
it is unknown how an outlier in a ranking affects the examination probability. To address this gap, we conduct an eye-tracking study.
We ask participants to interact with search engine result pages, as they normally would, and find the items that they prefer and think are relevant.
We track their eye movements via an online webcam-based eye tracking service.\footnote{RealEye, \url{https://realeye.io}.}
We use two scenarios, e-commerce, and scholarly search. 
We focus on a list view; in both scenarios, participants are able to see all items in one page.
For each scenario, we include two result pages, one without an outlier item and one with (as in Fig.~\ref{fig:eyetracking-smartphone-outlier}).
In the absence of outliers we expect participants to follow the position bias examination assumption~\citep{joachims2005accurately}; in the presence of outliers, we expect that users' attention is diverted towards them.

We recruit 40 university students and staff for both scenarios.
In the instructions, we describe the overall goal of the research and ask participants to read the instructions carefully. We describe what webcam-based eye tracking is and that the eye-tracking service will ask them for access to their webcam. We instruct participants to first read and understand the query and then start scanning the result page as if they submitted the query themselves.

\input{figures/tex_figures/DistributionCitations.tex}
\input{figures/tex_figures/NumberOutliers.tex}

For reporting, we consider four eye-tracking measures based on participants' eye fixations: 
%fixation count, time spent, average Time To First Fixation (TTFF), and revisit count. 
\begin{enumerate*}
\item fixation count (the number of fixations within an area; more fixation means more visual attention); 
\item time spent (shows the amount of time that participants spent on average looking at an area); 
\item Time To First Fixation (TTFF; the amount of time that it takes participants on average to look at one area for the first time); and \item revisit count (indicates how many times on average participants looked back at the area)~\citep{fiedler-2020-guideline}. 
\end{enumerate*}

\subheader{Outliers in e-commerce}
For this scenario, we mimick the Amazon Marketplace search result page.%\footnote{\url{https://www.amazon.com}}
Fig.~\ref{fig:eyetracking-smartphone-outlier} depicts our example ranked list with an outlier. The third item in the list stands out from other items for different reasons, including price and sales-related tags (e.g., being on sale), as well as other information that is available for this item. 
Comparing Fig.~\ref{fig:eyetracking-no-outlier-heatmap} and \ref{fig:eyetracking-smartphone-outlier-heatmap}, we see that in the presence of an outlier, items at the top of the list receive less attention, contradicting the position bias assumption.

\begin{figure}
    \vspace{-3.5mm}
    \begin{tabular}{@{}c@{~}c@{~}c@{}}
    \subfloat[]{%
          \includegraphics[,height=3.64cm,width=.32\linewidth]{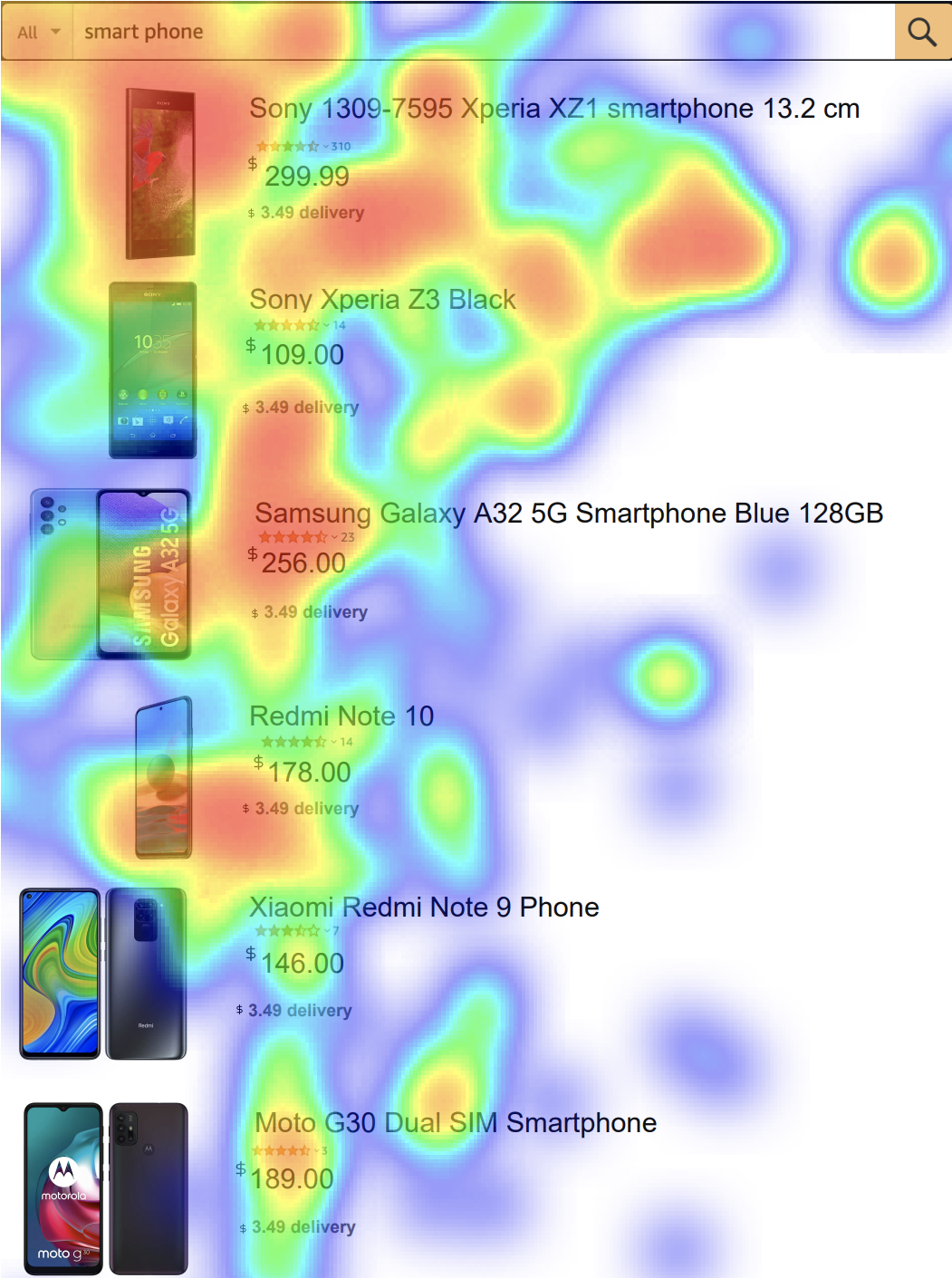}
          \label{fig:eyetracking-no-outlier-heatmap}}
     &
    \subfloat[]{%
          \includegraphics[height=3.64cm,width=.32\linewidth]{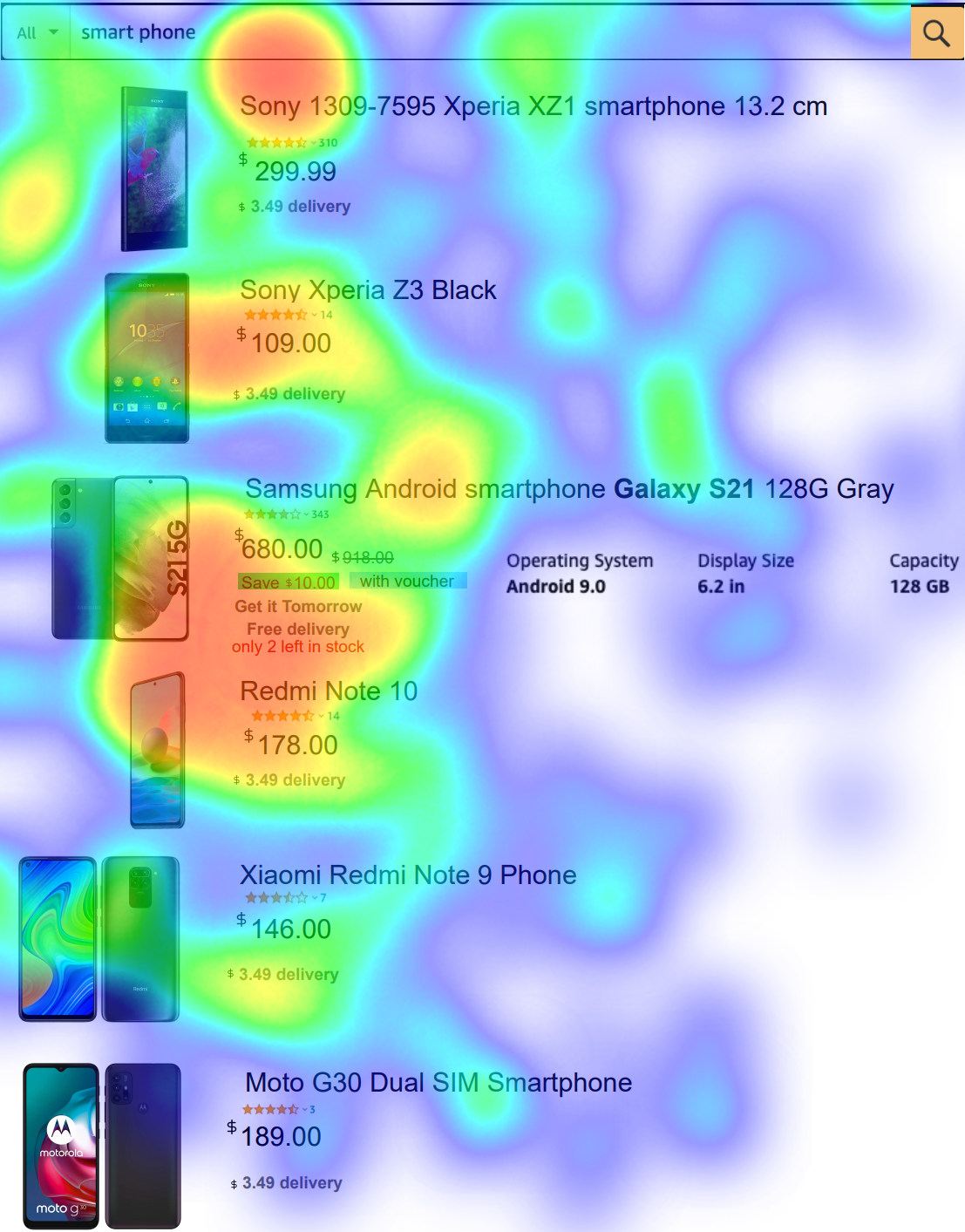}
          \label{fig:eyetracking-smartphone-outlier-heatmap}} 
    &
    \subfloat[]{%
          \includegraphics[height=3.64cm,width=.32\linewidth]{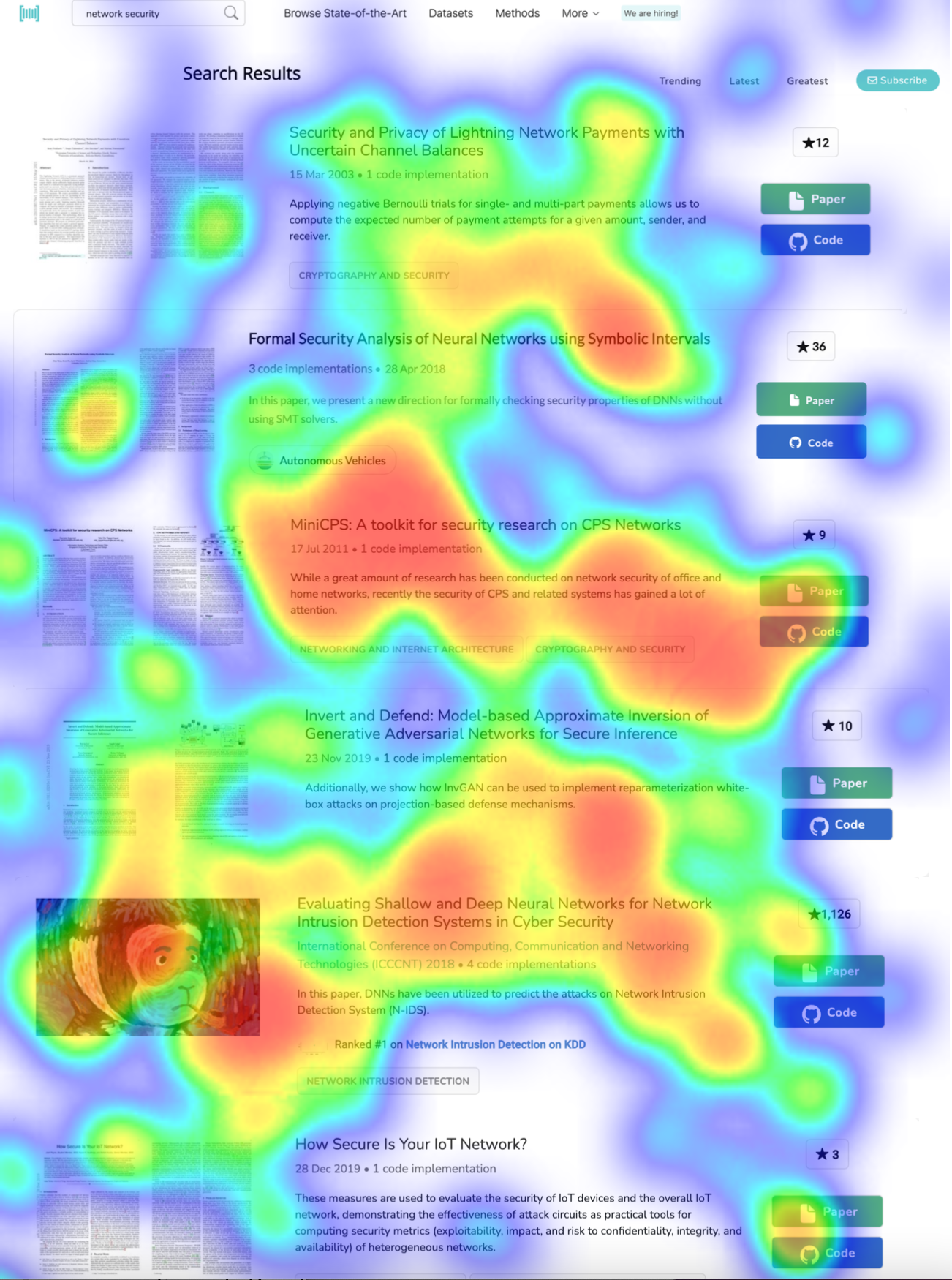}
          \label{fig:eyetracking-paper-outlier-heatmap}}
    \end{tabular}
\caption{Examples used in the eye-tracking study. 
%(a) and (b) concern an e-commerce example; (c) is a scholarly search example. 
(a) Heatmap  for a result list without outlier for the query ``smart phone''. Items at top ranks receive more attention, following the position bias assumption.
(b) Heatmap  for a similar page (the same list as in Fig.~\ref{fig:eyetracking-smartphone-outlier}) but with one outlier, at position 3. Participants exhibit increased attention towards and around the outlier item. 
(c) Heatmap for a scholarly search example, with an outlier (at position 4).
% with a different thumbnail image and large number of GitHub repository stars. 
%The outlier item draws lots of attention, and contradicts the cascade examination assumption.
}
%\label{fig:eyetracking}
  \end{figure}
 
Fig.~\ref{fig:eyetracking}(a) reports the eye-tracking measures for the e-commerce scenario.
We highlight the outlier in each list with an asterisk.
In the no-outlier condition, participants exhibit linear behavior in terms of scanning the items. The highest number of fixations, time spent, revisits belongs to the items on the top of the list, and it decreases as we go down the list. TTFF demonstrates a linear behavior of the first fixation time, that is, most of the participants started scanning the results from top to bottom. 
The ranked list with an outlier exhibits very different measurements. Attention is more focused on the outlier item. 
Also, we see that from the TTFF values, the average time for the first fixation is the lowest for this item, suggesting that the scanning order and exposure are influenced by the existence of the outlier. 
This is also evident by comparing the heatmaps in Fig.~\ref{fig:eyetracking-no-outlier-heatmap} and~\ref{fig:eyetracking-smartphone-outlier-heatmap}.

\if0
\begin{figure*}
\subfloat[][]{\includegraphics[height=3cm]{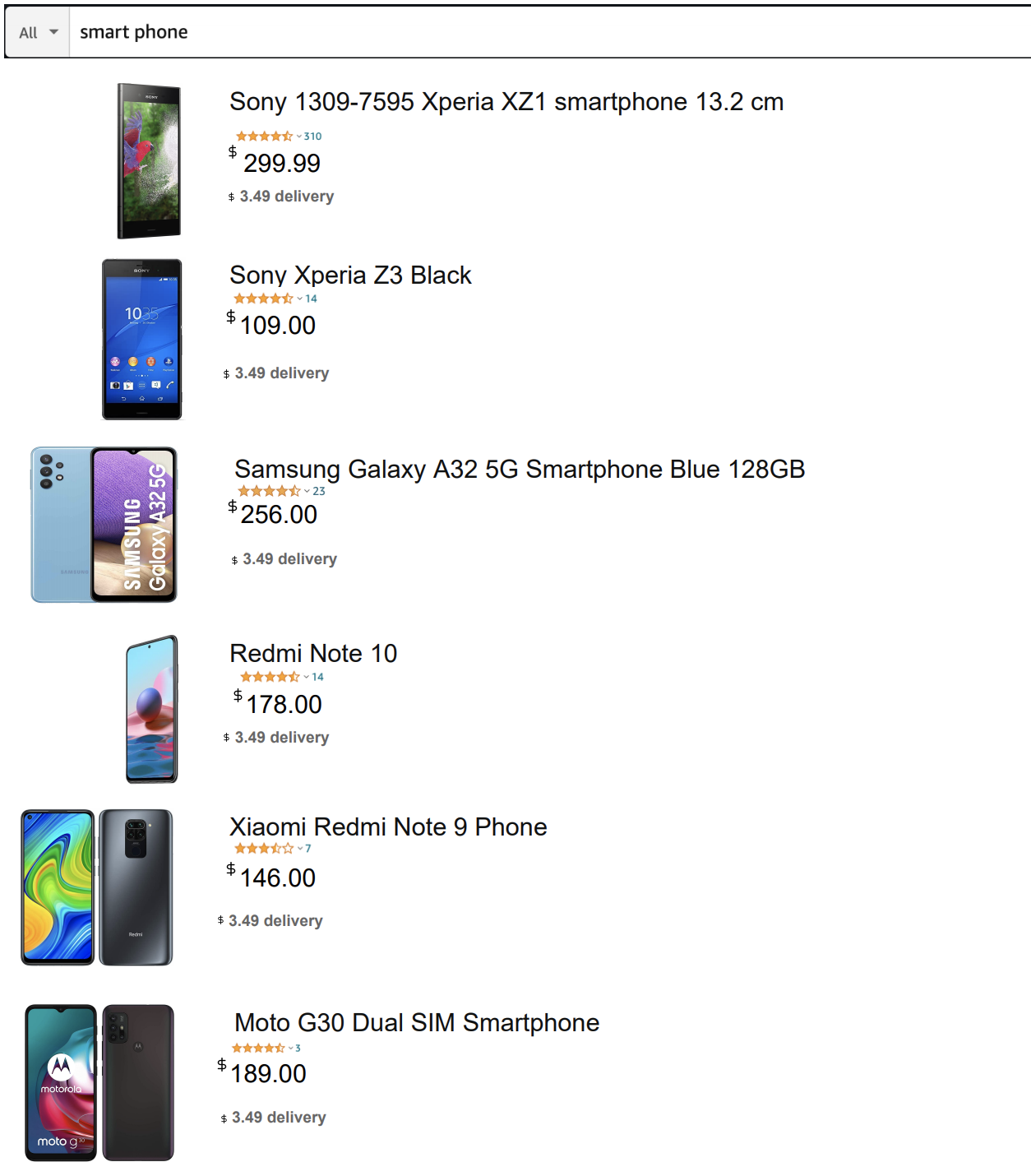}\label{fig:eyetracking-no-outlier}}~
\subfloat[][e-Commerce example with no outlier]{\includegraphics[height=3cm]{./figures/smartphone_no_outliers_heatmap.png}\label{fig:eyetracking-no-outlier-heatmap}}~
\subfloat[][]{\includegraphics[height=3cm]{./figures/smartphone_outlier.png}\label{fig:eyetracking-smartphone-outlier}}~
\subfloat[][e-Commerce example with outlier]{\includegraphics[height=3cm]{./figures/smartphone_outlier_heatmap.png}\label{fig:eyetracking-smartphone-outlier-heatmap}}~
\subfloat[][]{\includegraphics[height=3cm]{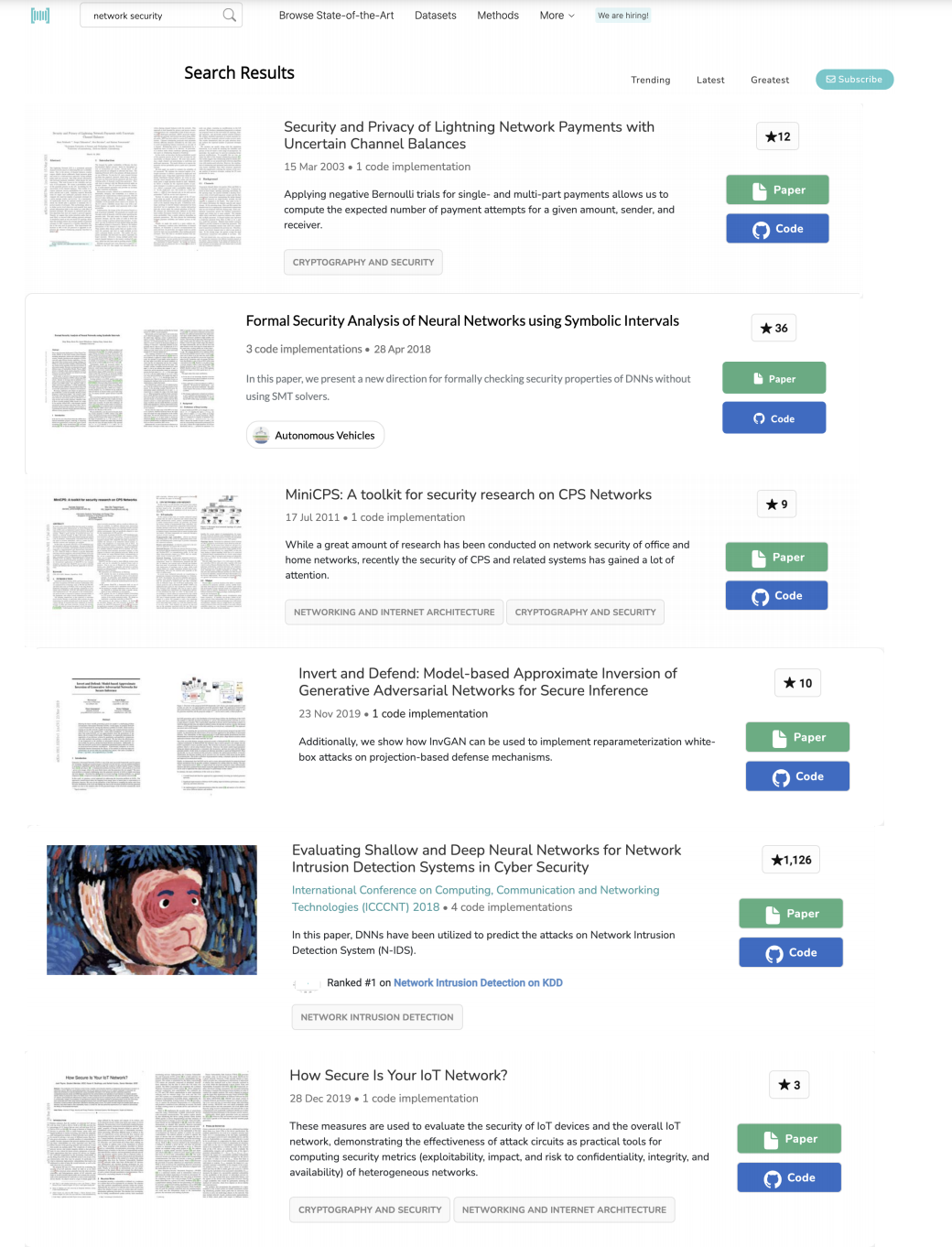}\label{fig:eyetracking-paper-outlier}}~
\subfloat[][Scholarly example with outlier]{\includegraphics[height=3cm]{./figures/paperswithcode_outlier_heatmap.png}\label{fig:eyetracking-paper-outlier-heatmap}}
\end{figure*}
\fi
\if0
 \begin{figure}             
\includegraphics[height=6.5cm,width=.35\linewidth]{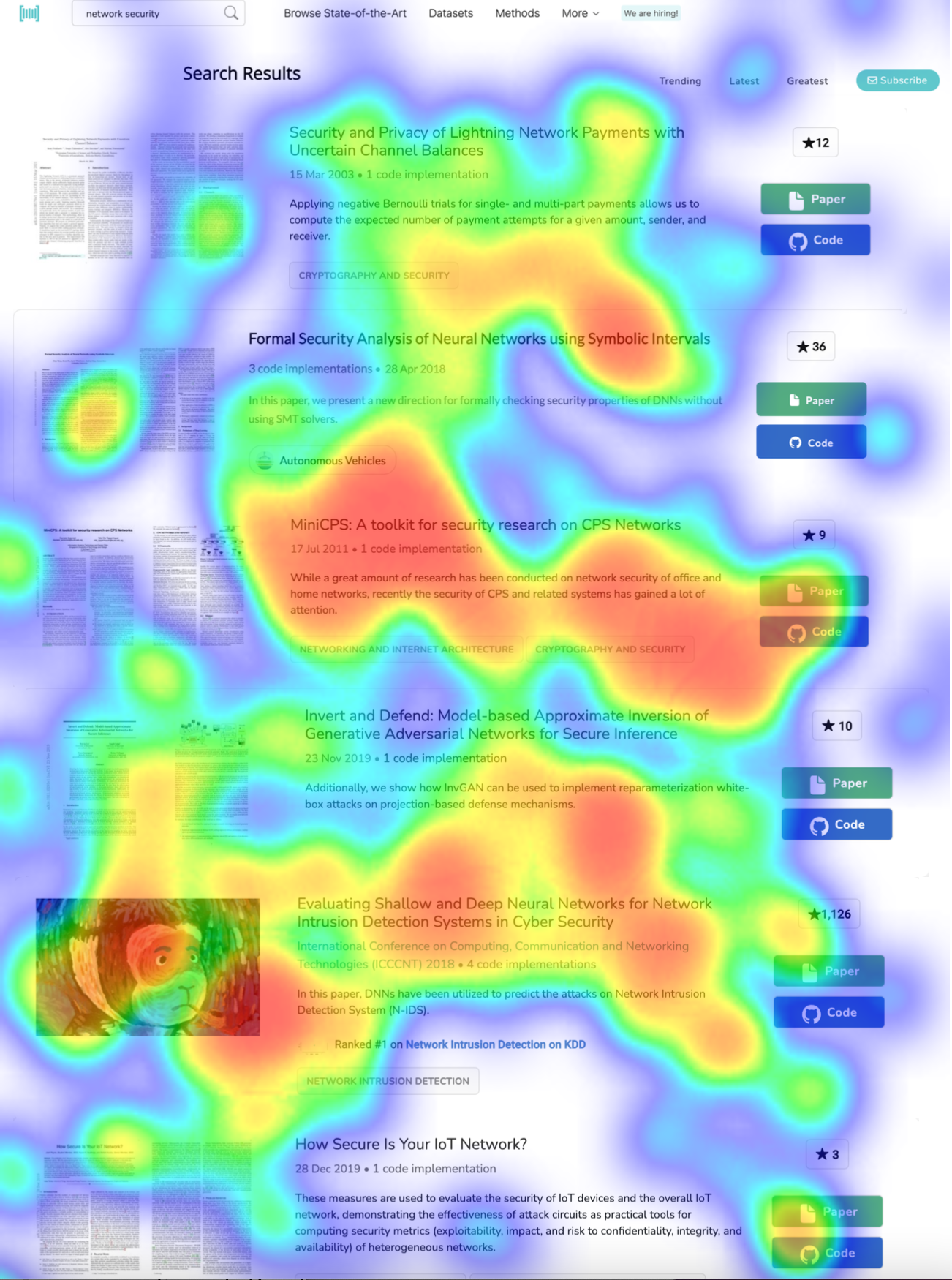}

\caption{Scholarly search example used in the eye-tracking study. The outlier is at the fourth position with its different thumbnail image and a large number of GitHub repository stars. The outlier item draws lots of attention and contradicts the cascade examination assumption.}
\label{fig:eyetracking-paper-outlier-heatmap}
\end{figure}
\fi

\pgfplotsset{
    discard if/.style 2 args={
        x filter/.code={
            \edef\tempa{\thisrow{#1}}
            \edef\tempb{#2}
            \ifx\tempa\tempb
                \def\pgfmathresult{inf}
            \fi
        }
    },
    discard if not/.style 2 args={
        x filter/.code={
            \edef\tempa{\thisrow{#1}}
            \edef\tempb{#2}
            \ifx\tempa\tempb
            \else
                \def\pgfmathresult{inf}
            \fi
        }
    }
}

\input{figures/tex_figures/EyeTracking.tex}

\subheader{Outliers in scholarly search}
In the second scenario, we study the effect of outliers on scholarly search result pages. To this end, we mimick the result page from PapersWithCode.\footnote{\url{https://paperswithcode.com}} 

To save space, we omit the eye fixation heatmap for the result page without outliers; it shows the familiar F-shape.
Fig.~\ref{fig:eyetracking-paper-outlier-heatmap} shows the eye fixation heatmap for the result page \emph{with} an outlier item; the fourth item has a different thumbnail and a large number of GitHub stars, making this item stand out in the list. 
Similar to the e-commerce scenario, the eye fixations are very different from the F-shape typical for the no-outlier case. 
Fixations and time spent are the highest for the outlier item, suggesting that it draws lots of attention, and contradicts the cascade examination assumption.

%\mdr{Say something about the measures reported in Figure~\ref{fig:eyetracking}(b).}
%relevant_outliers_at_each_position_MAD.csv
However, different from the e-commerce case, we do not observe as big a difference in the TTFF values between the conditions with and without outliers, suggesting that the order of item scans is not affected as much as in the e-commerce example. 
%\moh{we should also mention the case of items that fall lower that outlier in the ranking. they seem to receive very little attention, compared to the no-outlier case and the other items in the same ranking. this is another side-effect of having an outlier. }

%\mdr{I am not sure I understand the narrative structure here. We have several headers in this subsection, but then we have a separate subsection for the TREC data\ldots\ what's the point? Please make a clearer distinction between purpose; design; outcomes; conclusions, for each of the two setting (e-commerce and academic), and make sure that the section/subsection/heading structure follows these distinctions.}
\if0
\smallskip\noindent%
In this section, we have defined outliers in ranking, and we have seen that outliers occur in retrieval datasets and that user behavior is influenced by the presence of outliers. These findings motivate the definition of learning to rank model that is able to mitigate the presence of outliers in the top-$k$ positions for fair ranking.
\fi

%% file: figures/tex_figures/DistributionCitations.tex
%%% FIGURE %%% 

\begin{figure}[t]
\centering
\vspace*{2mm}
\newcommand{\SpacingX}{0.1em}
\newcommand{\Width}{0.5\textwidth}
\newcommand{\Height}{4cm}
\newcommand{\BarWidth}{0.001\textwidth}
\newcommand{\BarOffset}{0.00\textwidth}
\begin{tikzpicture}
    		\begin{axis}[
            symbolic x coords={0,1,2,3,4,5,6,7,8,9,10,11,12,13,14,15,16,17,18,19,20,21,22,23,24,25,26,27,28,29,30,31,32,33,34,35,36,37,38,39,40,41,42,43,44,45,46,47,48,49,50,51,52,53,54,55,56,57,58,59,60,61,62,63,64,65,66,67,68,69,70,71,72,73,74,75,76,77,78,79,80,81,82,83,84,85,86,87,88,89,90,91,92,93,94,95,96,97,98,99,100,101,102,103,104,105,106,107,108,109,110,111,112,113,114,115,116,117,118,119,120,121,122,123,124,125,126,127,128,129,130,131,132,133,134,135,136,137,138,139,140,141,142,143,144,145,146,147,148,149,150,151,152,153,154,155,156,157,158,159,160,161,162,163,164,165,166,167,168,169,170,171,172,173,174,175,176,177,178,179,180},
%            xtick=data, 
            %xtick style={draw = none},  %was not here
            xticklabel style = {font=\tiny,yshift=0.5ex},
            yticklabel = {
        \pgfmathparse{\tick/1000}
        \pgfmathprintnumber{\pgfmathresult}\,K
    },
            xticklabels = {,,},
            xlabel = Rankings,
            ymin=0,
            ylabel= \# Citations, 
            width=\Width,
            height=\Height,
            y label style={yshift=-1.1em, font=\small},
            x label style={yshift=1.3em, font=\small},
            legend style={fill opacity=0.75, draw opacity=1,text opacity=1},
            bar width=\BarWidth,
            enlarge x limits=0.015,
            enlarge y limits=0.003,
            mark size=0.6pt,
        	]       
        	\addplot[ybar,bar shift = 0*\BarOffset,fill=blue!40!white, color=blue!40!white] table[x=index, y=max,col sep=comma]{figure-data/citations_in_sessions.csv}; 
        	\addplot[ybar, bar shift = 0*\BarOffset,color=white,fill=white] table[x=index, y=min,col sep=comma]{figure-data/citations_in_sessions.csv}; 
        	\addplot[bar shift = 0*\BarOffset,draw=none, mark=*, color=blue!60!black, style={ultra thin}] table[x=index, y=mean,col sep=comma]{figure-data/citations_in_sessions.csv}; 

    		\end{axis}

	\end{tikzpicture}%
	
	\caption{Distribution of the number of citations of top-20 papers returned for test queries in the TREC Fair Ranking track 2020 data.}
%	 The variance between mean and maximum number of citations that can be seen in several rankings implies that the data is outlier-prone w.r.t. this feature.}
	\label{fig:citations-in-sessions}

\end{figure}

%% file: figures/tex_figures/NumberOutliers.tex
\begin{figure}[!t]
\centering

\newcommand{\SpacingX}{0.1em}
\newcommand{\Width}{0.5\textwidth}
\newcommand{\Height}{4cm}
\newcommand{\BarWidth}{0.005\textwidth}
\newcommand{\BarOffset}{0.005\textwidth}

\begin{tikzpicture}
    		\begin{axis}[
            ybar stacked,
            symbolic x coords={1,2,3,4,5,6,7,8,9,10,11,12,13,14,15,16,17,18,19,20},
            xtick=data, 
            %xtick style={draw = none},  %was not here
            xticklabel style = {font=\tiny,yshift=0.5ex},
            xlabel = Position,
            ymin=0,
            ylabel= \# Outliers, 
            width=\Width,
            height=\Height,
            y label style={yshift=-1.6em, font=\small},
            x label style={yshift=1.0em, font=\small},
            legend style={fill opacity=0.75, draw opacity=1,text opacity=1},
            bar width=\BarWidth,
            enlarge x limits=0.04,
        	]        
        	\addplot [bar shift = -\BarOffset,fill=orange!65!white] table[x=Position, y=Outliers,col sep=comma]{figure-data/non_relevant_outliers_at_each_position_MedKNN.csv}; \label{medknn_non}
        	\addplot [bar shift = -\BarOffset,fill=yellow!30!white] table[x=Position, y=Outliers,col sep=comma]{figure-data/relevant_outliers_at_each_position_MedKNN.csv};\label{medknn}
        	
    \makeatletter
	\pgfplots@stacked@isfirstplottrue
	\makeatother
	        	
       	\addplot [forget plot, bar shift = 0*\BarOffset,fill=blue!50!white] table[x=Position, y=Outliers,col sep=comma]{figure-data/non_relevant_outliers_at_each_position.csv};\label{copod_non}
        	\addplot [bar shift = 0*\BarOffset,fill=cyan!30!white] table[x=Position, y=Outliers,col sep=comma]{figure-data/relevant_outliers_at_each_position.csv};\label{copod}
	
	\makeatletter
	\pgfplots@stacked@isfirstplottrue
	\makeatother
	        	
       	\addplot [forget plot, bar shift = +\BarOffset,fill=asparagus!85!white] table[x=Position, y=Outliers,col sep=comma]{figure-data/non_relevant_outliers_at_each_position_MAD.csv};\label{mad_non}
        	\addplot [bar shift = +\BarOffset,fill=cambridgeblue!35!white] table[x=Position, y=Outliers,col sep=comma]{figure-data/relevant_outliers_at_each_position_MAD.csv};\label{mad}
%        	\legend{medknn non relevant, };
		\coordinate (legend) at (axis description cs:1.04,0.58);
    		\end{axis}
    	\node[draw,fill=white,inner sep=0pt,above left=0.5em] at (legend) {\small
    \begin{tabular}{lll}
%    non relevant & relevant \\
%    \ref{medknn_non} & \ref{medknn} & MedKnn\\
%    \ref{mad_non} & \ref{mad} & MAD\\
%    \ref{copod_non} & \ref{copod} & COPOD
%	& MedKnn & MAD & COPOD \\
%	non-rel. & \ref{medknn_non} & \ref{mad_non} & \ref{copod_non}\\
%	rel & \ref{medknn} & \ref{mad} & \ref{copod}\\
	MedKnn & \ref{medknn_non} non-rel.,  \ref{medknn} rel.\\
	COPOD & \ref{copod_non} non rel.,  \ref{copod} rel.\\
	MAD  & \ref{mad_non} non rel., \ref{mad} rel.
    \end{tabular}};
	\end{tikzpicture}%
	\caption{Number of outliers at each position w.r.t. different outlier detection methods, considering the number of citations as the observable feature. Each stacked bar shows the number of irrelevant and relevant outliers.}
	\label{fig:outliers_in_sessions}

\end{figure}

%% file: figures/tex_figures/EyeTracking.tex
\begin{figure*}
\newcommand{\SpacingX}{0.1em}
\newcommand{\Width}{0.29\textwidth}
\newcommand{\Height}{0.18\textwidth}
\newcommand{\BarWidth}{0.010\textwidth}
\newcommand{\BarOffset}{.006\textwidth}
\newcommand{\enlargelim}{0.15}
\centering
%\scalebox{.5}{\input{plot.tex}}
  \begin{tikzpicture}[font=\small]
    		\begin{axis}[
            ybar,
            symbolic x coords={1,2,3, 4,5,6},
            xtick=data, % was: xtick=data
            %xtick style={draw = none},  %was not here
%            xticklabels={,,,,,}, 
            xlabel = {},
%            xlabel = Position,
            ymin=0,
            ylabel= Fixation,
            width=\Width,
            height=\Height,
            y label style={yshift=-2.1em},
            x label style={yshift=.35em},
            legend style={fill opacity=0.75, draw opacity=1,text opacity=1,  at={(0.179\textwidth,0.1\textwidth)}},
            bar width=\BarWidth,
            enlarge x limits=\enlargelim,
        	]        
        	\addplot[bar shift = -\BarOffset,fill=blue!30!white] table[x=Position, y=No outlier,col sep=comma]{figure-data/fixation.csv};
        	\addplot[bar shift = \BarOffset, discard if={Position}{3},fill=red!30!white] table[x=Position,y=E-commerce w/ outlier,col sep=comma]{figure-data/fixation.csv};  
        	\addplot[bar shift=\BarOffset, discard if not={Position}{3}, fill=red!30!white,draw=black] table[x=Position,y=E-commerce w/ outlier,col sep=comma]{figure-data/fixation.csv};  
        	\draw (axis cs:3,80) node [xshift=0.4em] {\textasteriskcentered};
       	\legend{w/o outl., w outl.};
    		\end{axis}
	\end{tikzpicture}%
\hspace{\SpacingX}%
\begin{tikzpicture}[font=\small]
    		\begin{axis}[
            ybar,
            symbolic x coords={1,2,3, 4,5,6},
            xtick=data, 
            xlabel = {},
%            xlabel = Position,
            ymin=0,
            ylabel= Time spent (sec),
            width=\Width,
            height=\Height,
            y label style={yshift=-2.1em},
            x label style={yshift=.35em},
            legend style={fill opacity=0.75, draw opacity=1,text opacity=1},
            bar width=\BarWidth,
            enlarge x limits=\enlargelim,
        	]        
        	\addplot[bar shift = -\BarOffset,fill=blue!30!white] table[x=Position, y=No outlier,col sep=comma]{figure-data/timespentsec.csv};
        	\addplot[bar shift = \BarOffset, discard if={Position}{3},fill=red!30!white] table[x=Position,y=E-commerce w/ outlier,col sep=comma]{figure-data/timespentsec.csv};  
        	\addplot[bar shift=\BarOffset, discard if not={Position}{3}, fill=red!30!white,draw=black] table[x=Position,y=E-commerce w/ outlier,col sep=comma]{figure-data/timespentsec.csv};  
        	\draw (axis cs:3,1.3) node [xshift=0.4em] {\textasteriskcentered};
%        	\legend{without outliers, with outliers};
    		\end{axis}
	\end{tikzpicture}%
\hspace{\SpacingX}%
\begin{tikzpicture}[font=\small]
    		\begin{axis}[
            ybar,
            symbolic x coords={1,2,3, 4,5,6},
            xtick=data, % was: xtick=data
            %xtick style={draw = none},  %was not here
%            xticklabels={,,,,,}, 
            xlabel = {},
%            xlabel = Position,
            ymin=0,
            ylabel= TTFF (sec),
            width=\Width,
            height=\Height,
            y label style={yshift=-2.1em},
            x label style={yshift=.35em},
            legend style={fill opacity=0.75, draw opacity=1,text opacity=1,at={(0.02,0.98)},anchor=north west},
            bar width=\BarWidth,
            enlarge x limits=\enlargelim,
        	]        
        	\addplot[bar shift = -\BarOffset,fill=blue!30!white] table[x=Position, y=No outlier,col sep=comma]{figure-data/ttff.csv};
        	\addplot[bar shift = \BarOffset, discard if={Position}{3},fill=red!30!white] table[x=Position,y=E-commerce w/ outlier,col sep=comma]{figure-data/ttff.csv};  
        	\addplot[bar shift=\BarOffset, discard if not={Position}{3}, fill=red!30!white,draw=black] table[x=Position,y=E-commerce w/ outlier,col sep=comma]{figure-data/ttff.csv};  
        	\draw (axis cs:3,5) node [xshift=0.4em] {\textasteriskcentered};
%        	\legend{without outliers, with outliers};
    		\end{axis}
	\end{tikzpicture}%
\hspace{\SpacingX}%
\begin{tikzpicture}[font=\small]
    		\begin{axis}[
            ybar,
            symbolic x coords={1,2,3, 4,5,6},
            xtick=data, % was: xtick=data
            %xtick style={draw = none},  %was not here
%            xticklabels={,,,,,}, 
            xlabel = {},
%            xlabel = Position,
            ymin=0,
            ylabel= Revisits, 
            width=\Width,
            height=\Height,
            y label style={yshift=-2.1em},
            x label style={yshift=.35em},
            legend style={fill opacity=0.75, draw opacity=1,text opacity=1},
            bar width=\BarWidth,
            enlarge x limits=\enlargelim,
        	]        
        	\addplot[bar shift = -\BarOffset,fill=blue!30!white] table[x=Position, y=No outlier,col sep=comma]{figure-data/revisits.csv};
        	\addplot[bar shift = \BarOffset, discard if={Position}{3},fill=red!30!white] table[x=Position,y=E-commerce w/ outlier,col sep=comma]{figure-data/revisits.csv};  
        	\addplot[bar shift=\BarOffset, discard if not={Position}{3}, fill=red!30!white,draw=black] table[x=Position,y=E-commerce w/ outlier,col sep=comma]{figure-data/revisits.csv};  
        	\draw (axis cs:3,42) node [xshift=0.4em] {\textasteriskcentered};
%        	\legend{without outliers, with outliers};
    		\end{axis}
	\end{tikzpicture}%	
	
	\vspace*{-2mm}
	{(a) Eye tracking measurements for e-commerce search.}

  \begin{tikzpicture}[font=\small]
    		\begin{axis}[
            ybar,
            symbolic x coords={1,2,3, 4,5,6},
            xtick=data, % was: xtick=data
            %xtick style={draw = none},  %was not here
            xlabel = Position,
            ymin=0,
            ylabel= Fixation,
            width=\Width,
            height=\Height,
            y label style={yshift=-2.1em},
            x label style={yshift=.35em},
            legend style={fill opacity=1, draw opacity=1,text opacity=1,at={(0.0,1.35)},anchor=north west},
            bar width=\BarWidth,
            enlarge x limits=\enlargelim,
        	]        
        	\addplot[bar shift = -\BarOffset,fill=blue!30!white] table[x=Position, y=No outlier,col sep=comma]{figure-data/fixation.csv};
        	\addplot[bar shift = \BarOffset, discard if={Position}{4},fill=red!30!white] table[x=Position,y=Scholarly w/ outlier,col sep=comma]{figure-data/fixation.csv};  
        	\addplot[bar shift=\BarOffset, discard if not={Position}{4}, fill=red!30!white,draw=black] table[x=Position,y=Scholarly w/ outlier,col sep=comma]{figure-data/fixation.csv};  
        	\draw (axis cs:4,110) node [xshift=0.4em] {\textasteriskcentered};
        %	\legend{without outliers, with outliers};
    		\end{axis}
	\end{tikzpicture}%
\hspace{\SpacingX}%
\begin{tikzpicture}[font=\small]
    		\begin{axis}[
            ybar,
            symbolic x coords={1,2,3, 4,5,6},
            xtick=data, % was: xtick=data
            %xtick style={draw = none},  %was not here
            xlabel = Position,
            ymin=0,
            ylabel= Time spent (sec),
            width=\Width,
            height=\Height,
            y label style={yshift=-2.1em},
            x label style={yshift=.35em},
            legend style={fill opacity=0.75, draw opacity=1,text opacity=1},
            bar width=\BarWidth,
            enlarge x limits=\enlargelim,
        	]        
        	\addplot[bar shift = -\BarOffset,fill=blue!30!white] table[x=Position, y=No outlier,col sep=comma]{figure-data/timespentsec.csv};
        	\addplot[bar shift = \BarOffset, discard if={Position}{4},fill=red!30!white] table[x=Position,y=Scholarly w/ outlier,col sep=comma]{figure-data/timespentsec.csv};  
        	\addplot[bar shift=\BarOffset, discard if not={Position}{4}, fill=red!30!white,draw=black] table[x=Position,y=Scholarly w/ outlier,col sep=comma]{figure-data/timespentsec.csv};  
        	\draw (axis cs:4,2.35) node [xshift=0.4em] {\textasteriskcentered};
%        	\legend{without outliers, with outliers};
    		\end{axis}
	\end{tikzpicture}%
\hspace{\SpacingX}%
\begin{tikzpicture}[font=\small]
    		\begin{axis}[
            ybar,
            symbolic x coords={1,2,3, 4,5,6},
            xtick=data, % was: xtick=data
            %xtick style={draw = none},  %was not here
            xlabel = Position,
            ymin=0,
            ylabel= TTFF (sec),
            width=\Width,
            height=\Height,
            y label style={yshift=-2.1em},
            x label style={yshift=.35em},
            legend style={fill opacity=0.75, draw opacity=1,text opacity=1,at={(0.02,0.98)},anchor=north west},
            bar width=\BarWidth,
            enlarge x limits=\enlargelim,
        	]        
        	\addplot[bar shift = -\BarOffset,fill=blue!30!white] table[x=Position, y=No outlier,col sep=comma]{figure-data/ttff.csv};
        	\addplot[bar shift = \BarOffset, discard if={Position}{4},fill=red!30!white] table[x=Position,y=Scholarly w/ outlier,col sep=comma]{figure-data/ttff.csv};  
        	\addplot[bar shift=\BarOffset, discard if not={Position}{4}, fill=red!30!white,draw=black] table[x=Position,y=Scholarly w/ outlier,col sep=comma]{figure-data/ttff.csv};  
        	\draw (axis cs:4,10) node [xshift=0.4em] {\textasteriskcentered};
%        	\legend{without outliers, with outliers};
    		\end{axis}
	\end{tikzpicture}%
\hspace{\SpacingX}%
\begin{tikzpicture}[font=\small]
    		\begin{axis}[
            ybar,
            symbolic x coords={1,2,3, 4,5,6},
            xtick=data, % was: xtick=data
            %xtick style={draw = none},  %was not here
            xlabel = Position,
            ymin=0,
            ylabel= Revisits, 
            width=\Width,
            height=\Height,
            y label style={yshift=-2.1em},
            x label style={yshift=.35em},
            legend style={fill opacity=0.75, draw opacity=1,text opacity=1},
            bar width=\BarWidth,
            enlarge x limits=\enlargelim,
        	]        
        	\addplot[bar shift = -\BarOffset,fill=blue!30!white] table[x=Position, y=No outlier,col sep=comma]{figure-data/revisits.csv};
        	\addplot[bar shift = \BarOffset, discard if={Position}{4},fill=red!30!white] table[x=Position,y=Scholarly w/ outlier,col sep=comma]{figure-data/revisits.csv};  
        	\addplot[bar shift=\BarOffset, discard if not={Position}{4}, fill=red!30!white,draw=black] table[x=Position,y=Scholarly w/ outlier,col sep=comma]{figure-data/revisits.csv};  
        	\draw (axis cs:4,35) node [xshift=0.4em] {\textasteriskcentered};
%        	\legend{without outliers, with outliers};
    		\end{axis}
	\end{tikzpicture}%
	
	\vspace*{-2mm}
	{(b) Eye tracking measurements for scholarly search.}\\[2ex]

	\vspace*{-2mm}
  \caption{Eye tracking measurements for each position, based on participants’ eye fixations. The positions where outliers were shown are marked with an asterisk.}
  \label{fig:eyetracking}
	\vspace*{-4mm}
\end{figure*}
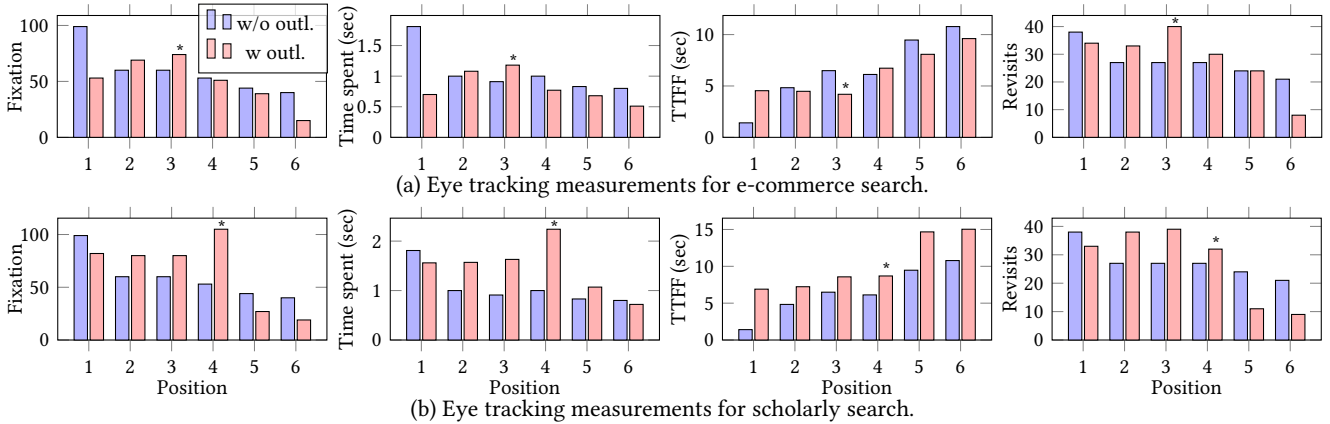

%% file: sections/04-method.tex
% !TEX root = ../main.tex

\vspace*{-2mm}
\section{Mitigating Outlierness in Fair Learning to Rank}
\label{sec:method}

We now present a ranking algorithm for mitigating outlierness for fairness in ranking, called \OurMethod, that simultaneously accounts for item fairness and outlierness effect requirements. 
Based on the observations reported in the previous section we know that outliers can influence exposure and examination order, in a way that can be considered as a type of bias. We take a first step towards mitigating the outlierness phenomenon by proposing a remarkably simple, yet effective solution that removes outliers from the top positions where the distribution of exposure is most critical. Our solution aims at decreasing outlierness in the \topk positions, while retaining the ranking's utility and fairness with position-based assumptions. 

\OurMethod{} is based on the linear programming method described in Section~\ref{sec:background}. In addition to optimizing for user utility while staying within the fairness constraints, our goal is to reduce the number of outliers in the \topk of all rankings.  
\OurMethod{} has two steps. In the first, we search for the marginal rank probability matrix that satisfies item group fairness by solving a linear program that optimizes both for user utility and fewer outliers in the \topk items with linear constraints for fairness. 
In the second, we derive a stochastic ranking policy from the solution of the linear programming method using the Birkhoff-von Neumann decomposition~\citep{birkhoff-1940-lattice}; cf.\ also~\citep{singh2018fairness}.

\header{Step 1: Computing MRP matrix}
Let $\mathcal D_q=\{d_1, \dots, d_N \}$ be a set of items that we aim to rank for a given query $q$. Each ranking $\ranking(\cdot|q)$ corresponds to some permutation matrix $P_{\ranking}$ that re-orders the elements of $\mathcal D_q$.\footnote{For simplicity, we interchangeably use $\ranking(\cdot)$ and $\ranking(\cdot|q)$, as well as $\mathcal D$  and $\mathcal D_q$.} As discussed in Section~\ref{sec:outliers:definition}, to determine which items are outliers, we use the domain specific observable item features $\ObservableFeature(d_1),\dots, \ObservableFeature(d_N)$ that potentially impact the user's perception of an item.
Using these characteristics as features, one can use any outlier detection method to determine which items should be considered outliers. 
The majority of outlier detection methods, including the ones we use (Section~\ref{sec:outliers:definition}), find outliers by calculating scores that indicate the degree of outlierness. This results in a list of outlierness values $\DegreeOfOutlierness{d_1}{\mathcal{D}},\dots, \DegreeOfOutlierness{d_{\contextSize}}{\mathcal{D}}$, where $\contextSize \le N$ is the size of the outlierness context that the algorithm takes into account while detecting the outliers.
%Since we are mostly interested in whether an items is perceived as an outlier or not 
We define a vector $\OutlierVector_\mathcal{D}$, that contains, for each document, the information whether it is an outlier with respect to the full ranked list:
\begin{equation}
\label{eq:outliervector}
\OutlierVector_\mathcal{D} = \{ o_i
\}_i
\quad \text{ with }  o_i = \begin{cases}
 \DegreeOfOutlierness{d_i}{\mathcal{D}},  & \text{if } d_i \text{ is outlier in $\mathcal{D}$}\\
  0, & \text{ else}. 
 \end{cases}
\end{equation}
We use $\OutlierVector$ and $\OutlierVector_\mathcal{D}$ interchangeably.  Note that we are considering outliers in the context of the full list, $\contextSize=N$, i.e., the outlier detection algorithm takes the whole ranked list as input. We assume that items that are perceived as outliers in this context are likely to be perceived as outliers when seen in the smaller context of the \topk items. Below, we show that this heuristic works well in practice. 

\OurMethod{} works by pushing outliers away from the \topk.  Let $P_{\ranking}$ be the permutation matrix corresponding to a ranking $\ranking$.  The amount of outlierness  that a ranking $\ranking$ places in the \topk are equal to
%The number of outliers that a ranking $\ranking$ places in the top-$k$ is equal to 
\begin{equation}
\label{eq:outlierness}
\text{Outlierness}_k^\mathcal{D}(\ranking | \OutlierDetection) = \sum_{i=0}^k (\OutlierVector^{T} P_{\ranking})_i = \OutlierVector^{T} P_{\ranking} \TopKOnes,
\end{equation}
where  $\TopKOnes=(1,\dots, 1,0,\dots, 0)$ is a vector containing 1 for the first $k$ positions and $0$ for all positions after that.  Similarly,  the expected outlierness, that is placed in the \topk by $\mathbf{P}$ is given by 
\begin{equation}
\label{eq:expected-outlierness}
\text{Outlierness}_k^\mathcal{D}(\mathbf{P}| \OutlierDetection) =  \OutlierVector^{T} \mathbf{P} \TopKOnes. 
\end{equation}
While optimizing for user utility we can use the expected outlierness to add an objective that will function as a regularization term, penalizing ranking policies with outliers in the \topk. 
We extend the linear programming approach described in Section~\ref{sec:background} to solve:
\begin{equation}
\label{eq:lp:omit}
\begin{aligned} 
\mathbf{P}	=  \operatorname{arg\,max}_{\mathbf{P}} & \mathbf{u}^{T} \mathbf{P} \mathbf{v}  - \OutlierVector^{T}\mathbf{P} \TopKOnes\\
\text{such that }				&  \mathbf{P}  \text{ is doubly-stochastic}\\
					&\mathbf{P} \text{ is fair}. 
\end{aligned}
\end{equation} 
%where $\lambda$ is a hyperparameter that helps us to trade off the importance of removing the outliers and user utility. 
%
For item fairness, we adopt the disparate treatment constraints as described in Section~\ref{sec:background}. 
Both terms of the optimization objective, and the constraints for fairness and finding a doubly stochastic matrix are linear in $N^2$ variables. 
Hence, the resulting linear program can be solved efficiently using standard optimization algorithms~\citep{singh2018fairness}.

% For item fairness, we adopt the disparate treatment constraints proposed in~\citep{singh2018fairness} to ensure that position-based exposure is distributed among item groups proportional to their merit (see Eq.~\ref{equation:disparity}). Following the work by ~\citet{singh2018fairness} we define \emph{merit} as relevance indicated by the average utility of items in each group:
%\begin{equation}
%U(G_k|q) = \frac{1}{|G_k|}\sum_{d_i\in G_k} r_{d_i, q} \quad G_k\in \mathcal G.
%\end{equation}

\header{Step 2: Constructing a stochastic ranking policy}
The solution to the linear program $\mathbf P$ is a matrix indicating the probability of showing each item at any position. To generate actual rankings, we need to derive a stochastic ranking policy $\StochasticRanking$ from $\mathbf P$ and sample rankings $\ranking$ to present to users. We follow~\citep{singh2018fairness} and use Birkhoff-von Neumann decomposition to compute $\StochasticRanking$, which decomposes the doubly stochastic matrix $\mathbf P$ into the convex sum of permutation matrices $P = \theta_1P_{\ranking_1}+\dots+\theta_nP_{\ranking_n}$, with $0\le\theta_i\le 1, \sum_i{\theta_i}=1$~\citep{birkhoff-1940-lattice}.  This results in at most $(N-1)^2+1$ rankings $\ranking_i$~\citep{marcus1959diagonals}, corresponding to $P_{\ranking_i}$, that are shown to the user with probability $\theta_i$, respectively.\footnote{We used the implementation from \url{https://github.com/jfinkels/birkhoff}.}
% elaborate on that

\header{\OurMethod{} model summary}
Algorithm~\ref{alg:omit} presents an overview of \OurMethod. \OurMethod takes as input the initial ranking $D_q$ (optimized for utility), outlier detection method $\OutlierDetection$, outlierness context size $\contextSize$, and the number of top items that we aim to remove outliers $k$. In line~\ref{alg1:line1}, $\OutlierVector_\mathcal{D}$ is created for a given outlier detection technique and outlierness context size, followed by line~\ref{alg1:line2} where we create the $\TopKOnes$ list that takes into account the top of the list that we aim to mitigate outlierness. In line~\ref{alg1:line3}, we solve the linear program that jointly solves the fairness and outlierness problem and pass the stochastic ranking in line~\ref{alg1:line4} to the BvN decomposition algorithm. Finally, we return the output of the BvN method as the output.

\vspace*{1mm}
\begin{algorithm}[t]
\caption{Outlier mitigation for fair ranking (\OurMethod)}
\label{alg:omit}
\begin{algorithmic}[1]
\Require $\mathcal D_q$, $\OutlierDetection$, $\contextSize$, $k$
\Ensure $\StochasticRanking$
\State Create $\OutlierVector_\mathcal{D}$ as Eq.~\ref{eq:outliervector} using $\mathcal D_q$ and $\OutlierDetection$ \label{alg1:line1}
\State $\TopKOnes \gets (h_1, \ldots, h_n)$ such that $h_i=1$ if $i\le k$ else $0$
\label{alg1:line2}
\State  \textbf{P} $\gets$  $\operatorname{arg\,max}_{\textbf{P}}   \textbf{u}^{T} \textbf{P} \textbf{v}  - \OutlierVector^{T}\textbf{P} \TopKOnes$ such that \textbf{P}  is doubly-stochastic and fair (Eq.~\ref{eq:lp:omit}) 
\label{alg1:line3}
\State $\StochasticRanking \gets $ BvN-Decomposition(\textbf{P})
\label{alg1:line4}
\State Return $\StochasticRanking$
\end{algorithmic}
\end{algorithm}

%% file: sections/05-experiments.tex
% !TEX root = ../main.tex

\vspace*{-2mm}
\section{Experimental Setup}
\label{sec:experiments}
%\label{sec:experiment_setup}
%We describe our research questions, data, evaluation metrics, and compared methods. 
%evaluate empirically how far our algorithm is able to mitigate outlierness without damaging ranking utility and item fairness. We also experiment with three different outlier detection methods employed in our model to study the sensitivity of the algorithm to the choice of outlier detection method and compare their results.
%\maria{This paragraph goes to empirical results.}
We target the following research questions:
\begin{enumerate*}[nosep,label=(RQ\arabic*),leftmargin=*]
	\item  How do different outlier detection methods affect \OurMethod{}'s performance?% in terms of utility, fairness, and outlierness?
	\item  How does our \OurMethod{} trade-off between utility, fairness, and outlierness, compared to baselines?
	\item  We adopt the constraints proposed in~\citep{singh2018fairness} (called FOE) to optimize a ranked list for fairness and utility through linear programming, as described in Section~\ref{sec:method}.  Given that \OurMethod{} adds additional constraints, is the overall linear program more effective when we treat the doubly stochastic matrix constraints as hard or soft constraints?
	\item  How does changing the context of detecting outliers affect \OurMethod{}'s outlierness improvement and utility?
	\item  How does changing $k$ affect \OurMethod{}'s outlierness improvement and utility in the \topk positions?
\end{enumerate*}

%\header{Baselines}
\header{Data} We use data from the TREC Fair Ranking 2019 and 2020 track (see Section~\ref{sec:outliers:outliers_in_TREC_data}). We make the group definitions over the two datasets consistent. 
As for the TREC 2019 data, we bin the original article groups into two classes. 
For the TREC 2020 data, we adopt the group definitions from the original data, that is, documents are assigned to two groups based on their authors' h-index.
Moreover, we follow the TREC setup to generate query sequences, leading to multiple occurrences of the same query, using the provided frequencies. Specifically, we evaluate on a query sequence of size $10,000$, including all the queries in the evaluation data. 

\header{Evaluation metrics}
We evaluate methods for fair learning to rank in the presence of outliers in terms of utility, item fairness, and outlierness. For utility and fairness, we use NDCG and dTR (see Eq.~\ref{eq:dtr}), respectively, as our metrics and report their expected values.
%Throughout the experiments we use the number of citations of each paper as the observable feature, and since a search engine result page is typically designed as a listing of size 10~\citep{kelly2015many}, we use $k=10$ in the experiments. 
%We then use this as feature $\ObservableFeature(d)$ to extract the scores $\DegreeOfOutlierness{d}{\mathcal{D}}$ as described in section \ref{sec:method}.  
To measure the expected outlierness of the policy $\mathbf{P}$ up to position $\contextSize$
%\maria{Why do we use $N$ here? Before $N$ is defined as the number of items to be ranked. I would use $n$.}
in the ranking, we use $\text{Outlierness}_\contextSize^\mathcal{D}(\mathbf{P}| \OutlierDetection)$ as defined in Eq.~\ref{eq:expected-outlierness}. Similarly we define the expected number of outliers up to position $\contextSize$ in ranking for policy $\mathbf{P}$ as
\begin{equation}
\label{eq:expected-outliercount}
\#\text{Outliers}_\contextSize^\mathcal{D}(\mathbf{P}| \OutlierDetection) =  \OutlierVector^{T}_b \mathbf{P} \TopKOnes, 
\end{equation}
where $\OutlierVector^{T}_b = \mathds{1}_{>0}(\OutlierVector^{T}) $ is the binarized version of $\OutlierVector^{T}$  where each outlier item is assigned $1$, and all the rest are assigned $0$.

\header{Compared methods}
To evaluate \OurMethod{}, we build several baseline methods, combining different options for each component of our model (initial ranking, fairness of exposure, outlier mitigation):
\begin{itemize}[leftmargin=*,nosep]
	\item \textbf{Initial ranking:} The initial ranking of all compared methods is generated using ListNet~\cite{cao07}. ListNet is a learning to rank model, optimizing for utility. We use it in our experiments to create initial ranked lists, $\mathcal D_q$, using the click data provided in the training set, with $30$ maximum epoch, and a validation ratio of $0.3$.
	\item \textbf{Fairness of exposure}: We use two variants of FOE~\citep{singh2018fairness} based on hard vs.~soft doubly stochastic matrix constraints, and call them FOE$^H$ and FOE$^S$, respectively.\footnote{We used the implementation from \url{https://github.com/MilkaLichtblau/BA_Laura}.}
	\item \textbf{Outlier mitigation}: We employ two outlier mitigation techniques, namely, \baseline and \OurMethod. \baseline removes all the outlier items detected by $\OutlierDetection$ from the ranking, while \OurMethod is our proposed outlier mitigation method as described in Section~\ref{sec:method}.
\end{itemize}

\noindent%
We specify methods as combinations of the three components mentioned above. E.g., ``ListNet + FOE$^H$ + \OurMethod'' uses the initial ranked list produced by ListNet, applying FOE fairness post-processing with hard constraints and the \OurMethod outlier mitigation model.

%To evaluate how the extra objective for outlierness mitigation affect user utility and group fairness, we compare our model with two state-of-the-art baselines: \emph{ListNet} is a learning to rank model, optimizing for utility and \emph{FOE}\footnote{We used the implementation from \url{https://github.com/MilkaLichtblau/BA_Laura}.} is an item fairness post processing algorithm proposed by~\citet{singh2018fairness} that optimizes a ranked list for fairness and utility through linear programming as described in Section~\ref{sec:method}. 
%We further develop another baseline that primitively removes outliers from the ranked list. We refer to this baseline as $\baseline$. Using this baseline we first remove all the outlier items detected by $\OutlierDetection$ from the list, and then apply \emph{FOE} to optimize the ranking for fairness and utility.
%We use ListNet throughout the experiments to create initial ranked lists using the click data provided in the training set, with $30$ maximum epoch, and validation ratio of $0.3$. We evaluate on a query sequence of size $10,000$, including all the queries in the evaluation data. 

%% file: sections/06-results.tex
% !TEX root = ../main.tex

\vspace*{-2mm}
\section{Empirical Results}
\label{sec:results}
\label{sec:results1}

%\header{RQ2. How do different outlier detection methods affect performance compared to each other in terms of utility, fairness and outlierness?}
%Among the three outlier detection methods of our experiments, COPOD leads to slightly higher NDCG compared to MedKNN and MAD for TREC 2020 data, however the difference is not significant.
\noindent\textbf{Effect of outlier detection method.}
We address \textbf{RQ1} by changing the outlier detection method, while keeping the other parts of the model fixed. Table~\ref{tabel:results1} reports the results of using three different outlier detection methods in \OurMethod. For comparison, we also report the results of ListNet without outlier mitigation (row \textbf{a}) and report the relative improvements. 
All three outlier detection methods effectively reduce outlierness compared to the baselines. COPOD achieves the best results by reducing outlierness by $80.3\%$ and $80.6\%$ in terms of Outlierness@10 on the TREC 2019 and 2020 data, respectively. In terms of dTR, COPOD outperforms MAD and MedKNN on both datasets where it increases dTR by $4.5\%$ on TREC 2020. 
We see no significant difference in the utility achieved by the methods.
%, one exception is NDCG@10 for TREC 2020 data where MAD improves utility $1.9\%$ more than COPOD.
Given that COPOD is parameter-free and scalable, we prefer it over the other two methods. For the remaining experiments, we choose COPOD as the outlier detection method.
% and report results only with this outlier detection method.

%\header{RQ1. How does our models trade-off between utility, fairness and outlierness, compared to baselines?}
\header{Utility, fairness, and outlierness trade-offs}
To answer \textbf{RQ2}, we turn to Table~\ref{tabel:results1}, which shows the results for \OurMethod{} and the baseline methods in terms of utility, fairness, and outlierness.
% We show the empirical results of the models described in Section~\ref{sec:experiments} in Table~\ref{tabel:results2}. 
Although ListNet is purely optimized for utility, it does not achieve the highest NDCG in all cases. This suggests that optimizing for fairness and outlierness could even improve the utility.
%However, the interesting observation is that it is followed by \OurMethod which exhibits higher performance compard to the fairness baseline. In particular, \OurMethod's performance in terms of NDCG@5 and NDCG@10 is $24.08\%$ and $19.64\%$, respectively. 
As shown in Fig.~\ref{fig:outliers_in_sessions} there is a high density of outliers among top positions that are mostly irrelevant. Therefore, when \OurMethod pushes these items to lower positions, it improves outlierness and utility measures simultaneously. 
Moreover, we see that mitigating outlierness does not cause any significant effect on dTR, showing that \OurMethod is capable of retaining position-based item fairness.
%
%Now that we showed \OurMethod does not damage fairness and utility, we evaluate its performance in terms of mitigating outlierness. 
Table~\ref{tabel:results2} shows that \OurMethod effectively decreases the number of outliers in top-$10$ positions by at most $83.49\%$ (and $82.93\%$) when used with $\mathrm{FOE}^S$ (row \textbf{g} in the table) on the TREC 2019 (and 2020) data. For the outlierness metrics, these values are $80.29\%$ and $80.66\%$. Referring back to our data analysis, we observed a high density of non-relevant outlier items at the top of the list, indicating the high possibility of user distraction towards these non-relevant items, as suggested by our eye-tracking study.  

    \begin{table}
    \caption{Comparing loss in fairness and utility, with gains in outlierness for different outlier detection methods on the TREC 2019 and 2020 Fair Ranking data. Models used: (a) $\mathrm{ListNet}$ and (b) $\mathrm{ListNet}+\mathrm{FOE}^S+\mathrm{\OurMethod}$. $\Delta$ values denote the percentage of relative improvement compared to (a). * refers to statistically significant improvements compared to (a) using a two-tailed paired t-test ($p < 0.05$).}
    \label{tabel:outlier_method_results}
    \label{tabel:results1}
    \setlength{\tabcolsep}{3.1pt}
    \centering
    \resizebox{1.0\columnwidth}{!}{%
    %  \begin{tabular}{lllSSSSSSSSS}
      \begin{tabular}{l@{}clccccccc}
        \toprule
        \multicolumn{2}{c}{} &
    	\multicolumn{1}{c}{} &
        \multicolumn{2}{c}{NDCG $\uparrow$} &
        \multicolumn{1}{c}{Fairness $\uparrow$} &
        \multicolumn{2}{c}{\# Outliers $\downarrow$} &
        \multicolumn{2}{c}{Outlierness $\downarrow$}
          \\
        \cmidrule(r){4-5}
        \cmidrule(r){6-6}
        \cmidrule(r){7-8}
        \cmidrule{9-10}
     
     &
        \multicolumn{1}{c}{Model} &   
        \multicolumn{1}{c}{Outl.} &
        \multicolumn{1}{c}{@5} &
        \multicolumn{1}{c}{@10} &
        \multicolumn{1}{c}{dTR} &
        \multicolumn{1}{c}{@10} &
        \multicolumn{1}{c}{$\Delta (\%)$} &
        \multicolumn{1}{c}{@10} &
        \multicolumn{1}{c}{$\Delta (\%)$} \\
        \midrule
         \multirow{6}{*}{\rotatebox[origin=c]{90}{\textbf{TREC 2019}}}  & 
            \multirow{3}{*}{(a)}    & COPOD  & 0.671 & 0.757 & 0.982 & 1.260 & $-$ & 0.873 & $-$ \\
             &                      & MedKNN & 0.671 & 0.757 & 0.982 & 0.507 & $-$ & 0.432 & $-$ \\
             &                      & MAD    & 0.671 & 0.757 & 0.982 & 0.811 & $-$ & 0.599 & $-$ \\
            % \cmidrule(r){2-10}
            % & \multirow{3}{*}{(b)} & COPOD  & 0.673 & 0.758 & 1.058 & 1.225 & & 0.852 & $-$ \\
            %  &                     & MedKNN & 0.673 & 0.758 & 0.944 & 0.520 & & 0.445 & $-$ \\
            %  &                     & MAD    & 0.673 & 0.758 & 0.944 & 0.780 & & 0.580 & $-$ \\
              \cmidrule(l){2-10}
            & \multirow{3}{*}{(b)} & COPOD & 	0.667  & 0.753  & \textbf{0.995}\rlap{*} & 0.208\rlap{*} & \textbf{83.49} & 0.172\rlap{*} & \textbf{80.29} \\
             & & MedKNN & 					    \textbf{0.671}  & 0.756  & 0.991 & 0.102\rlap{*} & 79.88 & 0.094\rlap{*} & 78.24 \\
             & & MAD & 						    \textbf{0.671}  & \textbf{0.757}  & 0.990 & 0.290\rlap{*} & 64.24 & 0.205\rlap{*} & 65.77 \\

          \midrule 
            
            \multirow{6}{*}{\rotatebox[origin=c]{90}{\textbf{TREC 2020}}}  & 
                \multirow{3}{*}{(a)}    & COPOD  & 0.240 & 0.356 & 0.267 & 1.043 & $-$ & 0.755 & $-$ \\
             &                          & MedKNN & 0.240 & 0.356 & 0.267 & 0.783 & $-$ & 0.602 & $-$ \\
             &                          & MAD    & 0.240 & 0.356 & 0.267 & 1.456 & $-$ & 0.779 & $-$ \\
            % \cmidrule(r){2-10}
            % & \multirow{3}{*}{(b)} & COPOD  & 0.241 & 0.357 & 0.766 & 1.046 & & 0.758 & $-$ \\
            % & &                      MedKNN & 0.241 & 0.357 & 0.766 & 0.766 & & 0.583 & $-$ \\
            % & &                      MAD    & 0.241 & 0.350 & 0.766 & 1.464 & & 0.780 & $-$ \\
              \cmidrule(l){2-10}
            & \multirow{3}{*}{(b)} & COPOD  & 0.240 & 0.366\rlap{*} & \textbf{0.279}\rlap{*} & 0.178\rlap{*} & \textbf{82.93} & 0.146\rlap{*} & \textbf{80.66} \\
             &                     & MedKNN & 0.239 & 0.361 & 0.263 & 0.160\rlap{*} & 79.56    & 0.133\rlap{*} & 77.90 \\
             &                     & MAD    & \textbf{0.242} & \textbf{0.372}\rlap{*} &  \textbf{0.278}\rlap{*} & 0.430\rlap{*} &       70.46    & 0.202\rlap{*} & 74.10 \\
        \bottomrule
        
      \end{tabular}%
      }
    \end{table}

%In this section we report our empirical results on  

\begin{table}[t]
\caption{Comparing loss in fairness and utility, with gains in outlierness for COPOD on the TREC 2019 and TREC 2020 Fair Ranking data. Models used: (a) $\mathrm{ListNet}$; (b) $\mathrm{ListNet}+\mathrm{FOE}^H$; (c) $\mathrm{ListNet}+\mathrm{FOE}^S$; (d) $\mathrm{ListNet}+\mathrm{FOE}^H+RO$; (e) $\mathrm{ListNet}+\mathrm{FOE}^S+RO$; (f) $\mathrm{ListNet}+\mathrm{FOE}^H+\mathrm{\OurMethod}$; (g) $\mathrm{ListNet}+\mathrm{FOE}^S+\mathrm{\OurMethod}$. $\Delta$ values denote the percentage of relative improvement compared to (a). 
Other conventions are the same as in Table~\ref{tabel:results1}.
%* refers to statistically significant improvements compared to (a) using the two tailed paired t-test ($p < 0.05$).
}
\label{tabel:results2}
\setlength{\tabcolsep}{3.1pt}
\centering
\resizebox{1.0\columnwidth}{!}{%
  \begin{tabular}{l@{}cccccccc}
    \toprule
    \multicolumn{2}{c}{} &
    \multicolumn{2}{c}{NDCG $\uparrow$} &
    \multicolumn{1}{c}{Fairness $\uparrow$} &
    \multicolumn{2}{c}{\# Outliers $\downarrow$} &
    \multicolumn{2}{c}{Outlierness $\downarrow$}
    \\
    \cmidrule(r){3-4}
    \cmidrule(r){5-5}
    \cmidrule(r){6-7}
    \cmidrule{8-9}
    &
    \multicolumn{1}{c}{Model} &
    \multicolumn{1}{c}{@5} &
    \multicolumn{1}{c}{@10} &
    \multicolumn{1}{c}{dTR} &
    \multicolumn{1}{c}{@10} &
    \multicolumn{1}{c}{$\Delta (\%)$} &
    \multicolumn{1}{c}{@10} &
    \multicolumn{1}{c}{$\Delta (\%)$} \\
    \midrule
      %\cmidrule(r){}
      %\multirow{3}{*}{AAAA} & 
      %COPOD
      
      \multirow{7}{*}{\rotatebox[origin=c]{90}{\textbf{TREC 2019}}} &
      (a) & 0.671 & 0.757 & 0.982 & 1.260 & $-$ & 0.873 & $-$   \\
    & (b) & 0.670 & 0.756 & 0.991 & 1.235 & $-$ & 0.870 & $-$   \\
    & (c) & \textbf{0.673} & \textbf{0.758} & 0.982 & 1.225 & $-$ & 0.852 & $-$   \\
    & (d) & 0.663 & 0.697 & \textbf{0.996}\rlap{*} & 1.114 & 11.58 & 0.861 & 1.37  \\
    & (e) & 0.667 & 0.700 & 0.965 & 1.072\rlap{*} & 14.92 & 0.834 & 4.47  \\
    & (f) & 0.672 & 0.757 & \textbf{0.996}\rlap{*} & 1.080\rlap{*} & 14.28 & 0.753\rlap{*} & 13.74 \\
    & (g) & 0.667 & 0.753 & 0.995\rlap{*} & 0.208\rlap{*} & \textbf{83.49} & 0.172\rlap{*} & \textbf{80.29} \\
      
    \midrule 
    
    \multirow{7}{*}{\rotatebox[origin=c]{90}{\textbf{TREC 2020}}} &
      (a) & 0.240 & 0.356 & 0.267 & 1.043 & $-$ & 0.755 & $-$     \\
    & (b) & 0.237 & 0.355 & 0.249 & 1.073 & $-$ & 0.780 & $-$     \\
    & (c) & 0.241 & 0.357 & 0.262 & 1.046 & $-$ & 0.758 & $-$     \\
    & (d) & \textbf{0.242} & 0.362 &  \textbf{0.293}\rlap{*} & 1.143 & $-$9.58\phantom & 0.811 & $-$7.41 \\
    & (e) & \textbf{0.242} & 0.362 & 0.269 & 1.148 & $-$10.06 & 0.817 & $-$8.21 \\
    & (f) & 0.237 & 0.359 & 0.282\rlap{*} & 0.885\rlap{*} & 15.15 & 0.645\rlap{*} & 14.56   \\
    & (g) & 0.240 & \textbf{0.366}\rlap{*} & 0.279\rlap{*} & 0.178\rlap{*} & \textbf{82.93} & 0.146\rlap{*} & \textbf{80.66}   \\
      %\midrule
    \bottomrule

  \end{tabular}%
 }
\end{table}

\header{Hard vs.~soft constraint}
We turn to \textbf{RQ3} and experiment with two variants of the FOE model, which differ in the constraint for generating a doubly stochastic matrix (see Eq.~\ref{eq:lp:omit}). This constraint is important since the BvN algorithm can guarantee to generate valid permutations only if the input is doubly stochastic. We observed that forcing the convex optimization to output such matrices can make the constraints too hard to satisfy even when only optimizing for fairness and utility. Hence, the algorithm cannot find an optimum solution for many of the queries. For example, $\mathrm{FOE}^H$ cannot find solutions for $47\%$, and $46\%$ of the queries on TREC 2019 and 2020  data, respectively. 
%These numbers are $46.7\%$ and $46\%$ for $\OurMethod^H$. 
We return the original ranking as the output when $\mathrm{FOE}^H$ does not find an optimum solution. To fix this problem we implemented the constraint for doubly stochastic matrix as a soft constraint and we check for validity of the permutation matrices generated by the decomposition algorithm. Table~\ref{tabel:results2} demonstrates the effectiveness of the soft variant of $\mathrm{FOE}$ (row \textbf{g} vs.~\textbf{f}). 

\header{Effect of $\contextSize$}
To answer \textbf{RQ4}, we examine the effect of the $\contextSize$ parameter, which determines the outlierness context size.
We change $\contextSize$ from $10$ to $40$ items while keeping other parameters fixed, and observing the behavior of the model. This is important since the outlierness of an item depends on its context, e.g., an item can be considered as an outlier in the top-10 items, but may not be an outlier in the top-20 items. The two left plots in Fig.~\ref{fig:RQ3} show the outlierness improvements (compared to ListNet) and utility in terms of NDCG@10 for different values of $\contextSize$ on the TREC 2019 and 2020 data. We see that Outlierness@10 improves for larger values of $\contextSize$, suggesting that determining outlierness of items in a bigger pool of items is more accurate and allows \OurMethod to mitigate the outliers in the top-10 positions more effectively.

\header{Effect of $k$}
To answer \textbf{RQ5}, we study the effect of changing $k$ when optimizing for mitigating outlierness in \topk positions.
%, while keeping other parameters fixed. 
We are interested in finding out how utility and outlierness are influenced when optimizing for mitigating outlierness for a longer list of top ranks ranging from $10$ to $40$. This mimics the cases where more items can be shown to a user, hence outliers in longer lists can be observed by the user. The left plots in Fig.~\ref{fig:RQ3} depict the results for both datasets. 
% Arezoo read this please and see if it should be 
% Notice that we report NDCG@10, since we care about utility at top positions in all scenarios. 
We observe that outlierness improvement drops for greater values of $k$ since it is more challenging for the algorithm to push all outliers to lower positions. 
Fig.~\ref{fig:outliers_in_sessions} shows that most outliers are located at top positions, so greater values of $k$ do not necessarily translate to more outliers. The changes in utility scores of the lists are marginal, with a $0.5\%$ increase and $1.6\%$ decrease for larger values of $k$ for the TREC 2019 and 2020 data, respectively. The difference in utility of the two datasets is due to the fact that there are more relevant outliers in TREC 2019 than in TREC 2020.

\input{figures/tex_figures/WindowSize_K_2020.tex}

%% file: figures/tex_figures/WindowSize_K_2020.tex
\begin{figure}
\centering

\newcommand{\SpacingX}{0.1em}
\newcommand{\Width}{0.16\textwidth}
\newcommand{\Height}{0.1\textwidth}
\newcommand{\BarWidth}{0.001\textwidth}
\newcommand{\BarOffset}{0.00\textwidth}

	\begin{tikzpicture}[font=\small]
    		\begin{axis}[
            	symbolic x coords={10,15,20,25,30,35,40},
            	xlabel = $\contextSize$,
            	width=\Width,
            	height=\Height,
            	y label style={yshift=-0.8em},
            	x label style={yshift=0.7em},
            	bar width=\BarWidth,
			axis y line*=left,
			scale only axis,
			ylabel={NDCG@10 \ref{pgfplots:plot1} },
			ymin=0.752,	% Update here for value smaller than minimum (of all values of both window size and k plot for outlierness)
			ymax=0.758, 	% Update here for value bigger than max of all values 
			ytick={   0.752, 0.754,0.756},			% Update here
			yticklabels={  0.752, 0.754,0.756},	% Update here
        	]       
        	\addplot[line width=0.4mm,color=red!40!white] table[x=windowsize, y=user_utility,col sep=comma]{figure-data/window_size_2019.csv}; 
        	\label{pgfplots:plot1}
    		\end{axis}
        	\begin{axis}[
            	symbolic x coords={10,15,20,25,30,35,40},
            	axis x line=none,
            	width=\Width,
            	height=\Height,
            	bar width=\BarWidth,
			axis y line*=right,
			scale only axis,
			ymin=-10,	% Update here for value smaller than minimum (of all values of both window size and k plot for outlierness)
			ymax=85,	% Update here for value bigger than max of all values 
			ytick={   0,20,40,60,80},			% Update here
			yticklabels={ ,,,, }
        	]       
        	\addplot[line width=0.4mm,color=blue!40!white] table[x=windowsize, y=outlierness,col sep=comma]{figure-data/window_size_2019.csv}; 
        	\label{pgfplots:plot2}
    		\end{axis}

	\end{tikzpicture}\begin{tikzpicture}[font=\small]
    		\begin{axis}[
            	symbolic x coords={10,15,20,25,30,35,40},
            	xlabel = $k$,
            	x label style={yshift=0.7em},
            	width=\Width,
            	height=\Height,
         	bar width=\BarWidth,
			axis y line*=left,
			scale only axis,		
			ymin= 0.752,	% Update here for value smaller than minimum (of all values of both window size and k plot for outlierness)
			ymax=0.758, 	% Update here for value bigger than max of all values 
			ytick={   0.752, 0.754,0.756},	
			yticklabels={  ,, },
        	]       
        	\addplot[line width=0.4mm,color=red!40!white] table[x=k, y=user_utility,col sep=comma]{figure-data/outlireness_k_2019.csv}; 
        	\label{pgfplots:plot1}
    		\end{axis}
        	\begin{axis}[
            	symbolic x coords={10,15,20,25,30,35,40},
            	axis x line=none,
            	width=\Width,
            	height=\Height,
%			axis y line*=right,  %HERE we are removing the arrow but fucking up the y-label
			axis y line=right,
            	y label style={yshift=0.8em},
            	x label style={yshift=0.7em},
			bar width=\BarWidth,
			scale only axis,
			ylabel={Out. Imp. (\%)  \ref{pgfplots:plot2} },
         	mark size=0.6pt,
			ymin=-10,	% Update here for value smaller than minimum (of all values of both window size and k plot for outlierness)
			ymax=85,	% Update here for value bigger than max of all values 
			ytick={   0,20,40,60,80},	
			yticklabels={  0, 20, 40,60,80},	% Update here
        	]       
        	\addplot[line width=0.4mm,color=blue!40!white] table[x=k, y=outlierness,col sep=comma]{figure-data/outlireness_k_2019.csv}; 
        	\label{pgfplots:plot2}
    		\end{axis}
    		
	\end{tikzpicture}%
	
	\vspace*{-2mm}
	{(a) TREC Fair Ranking Track 2019 data.}
	
	\begin{tikzpicture}[font=\small]
    		\begin{axis}[
            	symbolic x coords={10,15,20,25,30,35,40},
            	xlabel = $\contextSize$,
            	width=\Width,
            	height=\Height,
            	y label style={yshift=-0.8em},
            	x label style={yshift=0.7em},
            	bar width=\BarWidth,
			axis y line*=left,
			scale only axis,
			ylabel={NDCG@10 \ref{pgfplots:plot1} },
			ymin=0.355,	% Update here for value smaller than minimum (of all values of both window size and k plot for outlierness)
			ymax=0.370, 	% Update here for value bigger than max of all values 
			ytick={   0.355, 0.360,0.365 },	
			yticklabels={   0.355, 0.360,0.365 },% Update here
        	]       
        	\addplot[line width=0.4mm,color=red!40!white] table[x=windowsize, y=user_utility,col sep=comma]{figure-data/window_size_2020.csv}; 
        	\label{pgfplots:plot1}
    		\end{axis}
        	\begin{axis}[
            	symbolic x coords={10,15,20,25,30,35,40},
            	axis x line=none,
            	width=\Width,
            	height=\Height,
            	bar width=\BarWidth,
			axis y line*=right,
			scale only axis,
			ymin=-10,	% Update here for value smaller than minimum (of all values of both window size and k plot for outlierness)
			ymax=90,	% Update here for value bigger than max of all values 
			ytick={ 0,20,40,60,80},			% Update here
			yticklabels={ ,,,,}
        	]       
        	\addplot[line width=0.4mm,color=blue!40!white] table[x=windowsize, y=outlierness,col sep=comma]{figure-data/window_size_2020.csv}; 
        	\label{pgfplots:plot2}
    		\end{axis}

	\end{tikzpicture}\begin{tikzpicture}[font=\small]
    		\begin{axis}[
            	symbolic x coords={10,15,20,25,30,35,40},
            	xlabel = $k$,
            	x label style={yshift=0.7em},
            	width=\Width,
            	height=\Height,
         	bar width=\BarWidth,
			axis y line*=left,
			scale only axis,		
			ymin=0.355,	% Update here for value smaller than minimum (of all values of both window size and k plot for outlierness)
			ymax=.370, 	% Update here for value bigger than max of all values 
			ytick={   0.355, 0.360,0.365 },	
			yticklabels={  ,, },
        	]       
        	\addplot[line width=0.4mm,color=red!40!white] table[x=k, y=user_utility,col sep=comma]{figure-data/outlireness_k_2020.csv}; 
        	\label{pgfplots:plot1}
    		\end{axis}
        	\begin{axis}[
            	symbolic x coords={10,15,20,25,30,35,40},
            	axis x line=none,
            	width=\Width,
            	height=\Height,
%            	axis y line*=right, 	%HERE we are removing the arrow but fucking up the y-label
            	axis y line=right,
            	y label style={yshift=0.8em},
            	x label style={yshift=0.7em},
			bar width=\BarWidth,
			scale only axis,
			ylabel={Out. Imp. (\%)  \ref{pgfplots:plot2} },
         	mark size=0.6pt,
			ymin=-10,	% Update here for value smaller than minimum (of all values of both window size and k plot for outlierness)
			ymax=90,	% Update here for value bigger than max of all values 
			ytick={ 0,20,40,60,80},	
			yticklabels={ 0,20,40,60,80},		% Update here
        	]       
        	\addplot[line width=0.4mm,color=blue!40!white] table[x=k, y=outlierness,col sep=comma]{figure-data/outlireness_k_2020.csv}; 
        	\label{pgfplots:plot2}
    		\end{axis}

	\end{tikzpicture}%
	
	\vspace*{-2mm}
	{(b) TREC Fair Ranking Track 2020 data.}\\[2ex]

	\vspace*{-2mm}
\caption{NDCG@10 and outlierness@10 improvement percentage for different values of $k$ and different sizes of the window in which we detect the outliers. 
}
\if0
% leave the interpretation etc to the main text; only state what the plot is about:
Outlierness improvement drops for bigger values of $k$, but \todo{the difference in Utility of the list is marginal}. $k$ is optimal at the value where the two lines cross.  
Both metrics are improved by increasing the window size.} %The on the x-a-axis we see }
\fi
\label{fig:RQ3}
\end{figure}
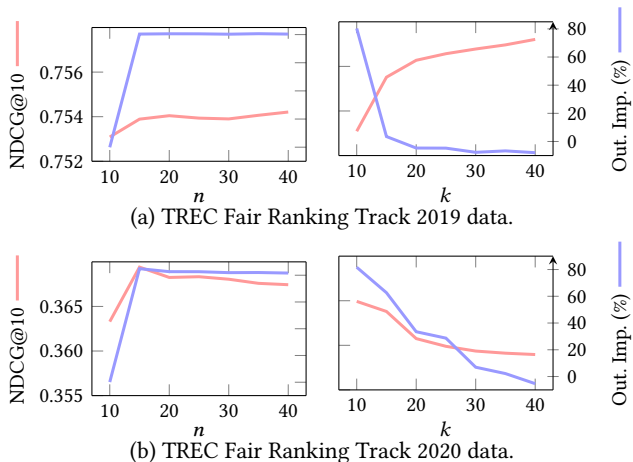

%% file: sections/07-related-work.tex
\vspace*{-2mm}
\section{Related Work}
\label{section:relatedwork}

\textbf{Bias in implicit feedback.}
Users' implicit feedback, such as clicks, can be a great source of relevance judgment that has been shown to help improve search quality~\citep{agarwal2019addressing}. clicks may be misleading due to different types of bias, which causes the probability of a click to differ from the probability of relevance.
Recent work on discovering and correcting for {different types of} bias in logged click data concerns position bias~\citep{joachims2005accurately, joachims2017unbiased}, presentation bias~\citep{yue2010beyond}, selection bias~\citep{ovaisi2020correcting}, trust bias~\citep{agarwal2019addressing, vardasbi2020inverse}, popularity bias~\citep{abdollahpouri2017controlling}, and recency bias~\citep{chen2019correcting}. 
Inter-item dependencies affect the perceived relevance of items~\citep{chuklin-2015-click}. 
We introduce a phenomenon that is anchored in inter-item dependencies and may result in biased clicks. Our work differs from previously discussed types of bias by considering inter-item relationships. 
Showing items as outliers can make them more attractive to users, influencing their perceived relevance. 
Presentation bias~\citep{yue2010beyond} concerns a related effect; bold keywords in titles make some items appear more attractive.  However, this definition of attractiveness is independent of other items in the list. 
\citet{metrikov2014whole} show that click-through rates can be manipulated by adding more images to top positions next to the ad slots on the search result page; they did not study the effect on item exposure or biased clicks.
We focus on the effect of outliers, as an inter-item dependency on the examination probability and item exposure, and introduce it as a potential source of bias in click data.

\header{Fair ranking}
Following \citep{zehlike2021fairness}, we distinguish two ways of measuring fairness of rankings.   
Work on \emph{probability-based} fairness determines the probability that a ranking is the result of a fair process~\citep{yang2017measuring, zehlike2017fa, asudeh2019designing, celis2020interventions, celis2017ranking, geyik2019fairness, stoyanovich2018online}.  
\emph{Exposure-based} methods determine the expected exposure for each item in the ranking and aim to ensure that this exposure is fairly distributed~\citep{biega2018equity, mehrotra2018towards,  morik2020controlling, sapiezynski2019quantifying, singh2018fairness, singh2019policy, yadav2019fair, diaz2020evaluating}.
Our work belongs to the second category.
To estimate the expected exposure of each item or group, we need to take into account different types of bias that the user might have when observing system outputs.  Previous work has mostly focused on position bias~\citep{biega2018equity,  singh2018fairness, yadav2019fair}.  We emphasize the role of inter-item dependencies in the exposure that an item receives, which can be a source of unfairness when not considered in computing the expected exposure.  We extend the re-ranking approach introduced in~\citep{singh2018fairness}, to not only produce fair rankings but also avoid showing the outliers in the top-$k$.  

%rank aware fairness 
% Considering the fairness definitions, there has been more exploration, ranging from group fairness to individual fairness \cite{} and from demographic parity to fairness of exposure \cite{singh2018fairness}.  
%TODO describe in a bit more detail how the different methods are approaching fair ranking? 
%In this work we argue that besides position bias there are other factors influencing the attention given to each item in the ranking. Based on our observations inter-item dependencies also play a role in the exposure that an item receives, which can be a source of unfairness when not considered in computing the expected exposure.  In this work we use an approach similar to \cite{singh2018fairness}, where we treat the problem as a convex optimization problem with linear constraints.  Additionally to the linear constraints that are in place to ensure fairness, we add a linear constraint for reducing the probability of showing an outlier in the top-$k$.  

An important effort for developing a benchmark for evaluating retrieval systems in terms of fairness is the \emph{TREC Fair Ranking track} (see footnote~\ref{footnote:FR}). 
%So far this track provided datasets for fair ranking extracted from the Semantic Scholar corpus for the task of fairly reranking academic paper abstracts. 
We expand the use of the track's resources to include the study of outliers in fair ranking.
%The objective of the track is to fairly represent relevant authors from several groups. 
%In this work we follow the same experimental setting and data as TREC fair ranking track.

\if0
 \header{Anomaly detection}
The problem of outlier detection is fundamental in data mining~\citep{wen-2006-ranking}. Various outlier detection methods have been proposed, including covariance based models~\citep{rousseeuw1999fast}, proximity based models like k-Nearest Neighbours Detector~\citep{ramaswamy2000efficient}, linear models like One-class SVM detector~\citep{scholkopf2001estimating}, ensemble based~\citep{zhao2019lscp}, neural network based~\citep{liu2019generative}, and Copula based models~\citep{li2020copod}.
These outlier detection methods can also be used to detect outliers in rankings. We are the first to use anomaly detection models to capture the influence of outlier items on fairness in ranking.
\fi

%% file: sections/08-conclusion.tex
% !TEX root = ../main.tex

\vspace*{-2mm}
\section{Conclusion \& Future Work}
\label{section:conclusion}

% Future work 
We introduced and studied a phenomenon related to fair ranking, that is, outlierness. 
We analyzed data from the TREC Fair Ranking track and found a significant number of outliers in the rankings. 
We hypothesized that the presence of outliers in a ranking impacts the way users scan the result page. We confirmed this hypothesis with an eye-tracking study in two scenarios: e-commerce and scholarly search.
We proposed \OurMethod{}, an approach to mitigate the existence and effect of outliers in a ranked list. 
With \OurMethod{}, we introduced a ranking constraint based on the outlierness of items in a ranking and combined it with fairness constraints. We formulated the problem of outlier mitigation as a linear programming problem and produced stochastic rankings, given an initial ranking.
Using \OurMethod{} one can reduce outliers in rankings without compromising user utility or position-based item fairness.  We analyzed the effects of different outlier detection approaches and compared their results. Our experiments also showed that there is a trade-off between the depth of outlier detection and user utility.
Now that we have established the impact of outliers in rankings, future work on fair ranking should consider the presence of outliers by default. 

One limitation of our work is our focus on removing the bold outliers defined in the context of the whole list from the \topk positions. We plan to improve our model to mitigate the outlierness of all sliding windows of size $k$. 
%list-based views and on clicks. Future work should consider other presentation formats and other aspects of user behavior (such as scrolling and next-page clicks).
In addition, we want to improve the performance for cases where outliers are relevant items, e.g., by 
%As we saw in this work, most outlier items were irrelevant to the user query, hence a linear constraint on the convex optimization led to high performance. We plan to investigate other techniques such that a relevant outlier item can be grouped with other items of the same outlier feature, thus not being an outlier anymore. 
considering alternative methods of outlier mitigation.
%, such as moving it to the top of the list, so that the attention it receives overlaps with the one of the position bias.
%A question that remains open is on the size of the context that users perceive outliers in. In this work we only consider outliers in the top-$k$ and simply assume that the user perceives outliers in relation to the other items that appear in the top-$k$ with it.  Insights whether items are perceived in the context of the result page, or a smaller or bigger window or rather with respect to some sliding window, might change the method we need to use to remove outliers from the lists and would hence contribute a lot to the solution of this problem. 
%
%TODO how specific can we even be? 
%
Finally, a natural extension of our work could be to quantify the effect that an outlier in the presentation of the ranking can have on the examination probability distribution;
%, either through a large scale user study or the careful analysis of user logs.  With the help of a good user model that captures the outlier effect, 
this could open the door to unbiased learning to rank approaches that counter the outlier bias in logged user data.  
%Moreover such an estimation of the examination probability could be directly used to calculate the fairness metrics, potentially making methods that avoid outliers redundant. 